\documentclass[%
reprint,
superscriptaddress,
 amsmath,amssymb,
 aps,
 prx
]{revtex4-2}

\usepackage{color,soul}
\sethlcolor{yellow}

\usepackage{graphicx}
\graphicspath{{figures/}}
\usepackage{dcolumn}
\usepackage{bm}
\usepackage[mathlines]{lineno}

\begin{document}

\preprint{APS/123-QED}

\title{Phase Randomness in a Semiconductor Laser:\\the Issue of Quantum Random Number Generation}

\author{Roman Shakhovoy}
\email{r.shakhovoy@goqrate.com}
\affiliation{QRate, Skolkovo, Russia}
\affiliation{NTI Center for Quantum Communications, National University of Science and Technology MISiS, Moscow, Russia}
\affiliation{Moscow Technical University of Communications and Informatics, Moscow, Russia}

\author{Marius Puplauskis}
\affiliation{Russian Quantum Center, Skolkovo, Russia}

\author{Violetta Sharoglazova}
\affiliation{QRate, Skolkovo, Russia}

\author{Alexander Duplinskiy}
\affiliation{QRate, Skolkovo, Russia}
\affiliation{NTI Center for Quantum Communications, National University of Science and Technology MISiS, Moscow, Russia}
\affiliation{Russian Quantum Center, Skolkovo, Russia}

\author{Denis Sych}
\affiliation{P.N. Lebedev Physical Institute of the Russian Academy of Science, Moscow, Russia}

\author{Elizaveta Maksimova}
\affiliation{QRate, Skolkovo, Russia}

\author{Selbi Hydyrova}
\affiliation{Russian Quantum Center, Skolkovo, Russia}
\affiliation{Bauman Moscow State Technical University, Moscow, Russia}

\author{Alexander Tumachek}
\affiliation{Moscow Technical University of Communications and Informatics, Moscow, Russia}

\author{Yury Mironov}
\affiliation{Moscow Technical University of Communications and Informatics, Moscow, Russia}

\author{Vadim Kovalyuk}
\affiliation{NTI Center for Quantum Communications, National University of Science and Technology MISiS, Moscow, Russia}
\affiliation{National Research University Higher School of Economics, Moscow, Russia}

\author{Alexey Prokhodtsov}
\affiliation{National Research University Higher School of Economics, Moscow, Russia}

\author{Grigory Goltsman}
\affiliation{Russian Quantum Center, Skolkovo, Russia}
\affiliation{National Research University Higher School of Economics, Moscow, Russia}

\author{Yury Kurochkin}
\affiliation{Quantum Research Centre, Technology Innovation Institute, Abu Dhabi, United Arab Emirates}

\date{\today}

\begin{abstract}
Gain-switched lasers are in demand in numerous quantum applications, particularly, in systems of quantum key distribution and in various optical quantum random number generators. The reason for this popularity is natural phase randomization between gain-switched laser pulses. The idea of such randomization has become so familiar that most authors use it without regard to the features of the laser operation mode they use. However, at high repetition rates of laser pulses or when pulses are generated at a bias current close to the threshold, the phase randomization condition may be violated. This paper describes theoretical and experimental methods for estimating the degree of phase randomization in a gain-switched laser. We consider in detail different situations of laser pulse interference and show that the interference signal remains quantum in nature even in the presence of classical phase drift in the interferometer provided that the phase diffusion in a laser is efficient enough. Moreover, we formulate the relationship between the previously introduced quantum reduction factor and the leftover hash lemma. Using this relationship, we develop a method to estimate the quantum noise contribution to the interference signal in the presence of phase correlations. Finally, we introduce a simple experimental method based on the analysis of statistical interference fringes, providing more detailed information about the probabilistic properties of laser pulse interference.

\end{abstract}

\maketitle


\section{Introduction}\label{sec:introduction}
Phase randomness between pulses of a gain-switched semiconductor laser is an essential ingredient of quantum key distribution (QKD) systems and quantum random number generators (QRNGs). Inasmuch as amplified spontaneous emission dominates below threshold in a semiconductor laser \cite{Petermann,Svelto}, phase correlations of the electromagnetic field are destroyed very quickly between laser pulses under the gain switching. Therefore, many authors often assume implicitly that pulses from a gain-switched laser have no phase relationship to each other. The security analysis of QKD protocols, particularly decoy-BB84 protocol \cite{Hwang2003,Lo2005,Wang2005}, assumes that the laser emits light pulses that are a mixture of coherent states with uniformly distributed phases. Similar assumption is usually made when considering laser pulse interference as a quantum entropy source for some optical QRNGs \cite{Jofre2011, Abellan2014, Yuan2014, Abellan2015, Marangon2018, Zhou2019, Shakhovoy2020}. In real experiments, however, phase correlations may still occur and may thus lead to loss of security. Therefore, phase diffusion between laser pulses should be treated carefully in these applications.

The main reason for correlations between phases of pulses emitted by a gain-switched semiconductor laser is an insufficient delay between subsequent pulses, during which the phase does not have enough time to “diffuse”. Also, a high value of the bias current does not allow the attenuation between laser pulses to be high enough to provide fast decoherence. It was estimated in \cite{Jofre2011} that enough for application in QRNG randomness could be achieved with the laser pulse repetition rate up to 20 GHz in assumption that attenuation between pulses reaches 100 dB. The same authors demonstrated later an optical QRNG with the distributed feedback (DFB) gain-switched laser operating at pulse repetition rate of 5.825 GHz \cite{Abellan2014}. In \cite{Kobayashi2014}, the phase randomness between pulses with the repetition rate of 10 GHz was demonstrated to be enough for QKD applications. 

The issue of the phase randomness of attenuated laser pulses used as quantum states for QKD has been widely discussed in the literature \cite{Kobayashi2014, Lo2007, Sun2012, Tang2013}. It was shown that phase correlations enhance the distinguishability of quantum states \cite{Lo2007}, which is directly related to the security of the decoy-BB84 protocol; therefore, when the phase between coherent states is not well randomized, the performance of the QKD system will be substantially reduced. In fact, it was demonstrated experimentally that Eve could employ phase correlations to compromise the quantum key \cite{Sun2012, Tang2013}. The dependence of the state distinguishability (imbalance of the quantum coin \cite{Gottesman2004}) on the degree of phase correlations between laser pulses or rather on the value of the standard deviation $\sigma_\varphi$  of phase fluctuations has been investigated numerically in \cite{Kobayashi2014}. Authors demonstrated that the imbalance decreases as the standard deviation increases converging to finite values for large  $\sigma_\varphi$.

In \cite{Shakhovoy2020}, there has been introduced the concept of the quantum reduction factor, which allows estimating the contribution of classical noise to the interference of laser pulses. The proposed approach assumes that $\sigma_\varphi$  is large enough ($\sigma_\varphi>2\pi$), such that one may consider any classical contribution to phase fluctuations to be buried under the noise of spontaneous emission. The situation $\sigma_\varphi<2\pi$ has not been considered in \cite{Shakhovoy2020}; however, the definition of the quantum reduction factor should be now modified to take the lack of phase randomness into account. In this paper, we propose a method, which allows including these correlations into the model for a QRNG based on the interference of laser pulses. We discuss this in section \ref{sec:improvingRandomness}.

Another motivation for this paper was that there is no well-established theoretical approach in the literature that would allow estimating the pulse repetition rate and appropriate values of the pump current, at which the phase randomness could be still considered purely quantum and thus sufficient for application in a QRNG. Here, we intend to formulate a theoretical approach, which provides a clear criterion for the phase diffusion efficiency at any pulse repetition rate and any value of the pump current.

In addition to a theoretical model, which allows analyzing the dependence of $\sigma_\varphi$ on laser parameters, it is useful to have a reliable experimental method to measure the phase diffusion between pulses of a gain-switched laser.
A standard approach is to use visibility of interference fringes as a criterion of phase randomness (see, e.g., \cite{Kobayashi2014}). Such an approach provides a useful experimental probe to estimate the effectiveness of the phase diffusion. However, it does not allow distinguishing various noise contributions to the interference signal, such as the photodetector noise or intrinsic fluctuations of laser pulse intensity.
We will demonstrate here that it is possible to set up the experiment in such a way that the experimental data would be sensitive to the mentioned classical noises, so that they could be separated from the phase fluctuations by choosing an appropriate mathematical model. We discuss this in section \ref{sec:discussion} (for more details see \cite{Shakhovoy2020}).

In section \ref{sec:theory}, we consider the most essential features of the interference of laser pulses with random phases, provide a method to estimate the quantum noise contribution to the interference signal in the presence of phase correlations, and develop a numerical approach to follow the dependence of the phase diffusion on the pump current and the pulse repetition rate. In section \ref{sec:experiment}, we provide the reader with experimental results on the phase diffusion measurements. Finally, in sections \ref{sec:discussion} and \ref{sec:conclusion} discussions and conclusions are given.

\section{Theoretical considerations}\label{sec:theory}
\subsection{Interference of laser pulses with random phases}
In most textbooks on optics, the interference of light in a Mach-Zehnder interferometer is usually considered using the example of a plane-polarized monochromatic wave, which is divided by a first beam splitter and gets into the interferometer arms. In the general case, one of the arms can be longer or, e.g., may contain some object. After passing through the interferometer arms, the two resulting waves are then met at the second beam splitter, where the interference occurs. The result of the interference depends on the phase acquired by the wave in the long arm of the interferometer (or rather by the phase difference between the arms). In the case of a quasi-monochromatic wave, the result of the interference will also depend on the relationship between the time delay $\Delta T$  of the long arm and the coherence time $\tau_c$  of light. If $\Delta T\ll\tau_c$ , one may assume that electric fields meeting at the second beam splitter have the form:
\begin{equation}\label{eq:fieldsInTermsOfE0}
\begin{split}
{E_1}(t)& = {E_0}\exp \left( {i{\varphi _0} + i{\omega _0}t} \right),\\
{E_2}(t) &= {E_0}\exp \left[ {i{\varphi _0} + i{\omega _0}(t + \Delta T)} \right],
\end{split}
\end{equation}
where  $\omega_0$ is the mid-frequency of the field and where we assumed for simplicity that both beam splitters of the interferometer are ideal 50:50 beam splitters. We also assume that losses in the interferometer arms are the same (or may be neglected), so that the real amplitudes of the interfering fields are equal to $E_0$. The result of the interference $S$ in one of the output ports of the interferometer will then be written as follows:
\begin{equation}\label{eq:resultOfInterference1}
S \equiv {\left| {{E_1} + {E_2}} \right|^2} = 4E_0^2{\cos ^2}\left( {\frac{{{\omega _0}\Delta T}}{2}} \right).
\end{equation}
It is clear from Eq.~\eqref{eq:resultOfInterference1} that the interference depends on the carrier frequency   of the field, which is a standard result.

A somewhat different result is obtained when considering the interference of the two monochromatic waves from independent laser sources \cite{Magyar63}. To observe the interference in this case, it is sufficient to take a single beam splitter and to send laser beams into its input ports. If the intensities of the two laser beams are the same, then the monochromatic waves interfering at the beam splitter can be written as follows:
\begin{equation}\label{eq:fieldsE1E2}
\begin{split}
{E_1}(t)& = {E_0}\exp \left( {i\varphi _1^0 + i{\omega _0}t} \right),\\
{E_2}(t)& = {E_0}\exp \left[ {i\varphi _2^0 + i{\omega _0}(t + \tau )} \right],
\end{split}
\end{equation}
where $\varphi_1^0$  and $\varphi_2^0$  are random initial phases of the fields and $\omega_0\tau$  corresponds to a phase difference due to the optical path difference of the light beams. The result of the interference in one of the output ports of the beam splitter can be written as follows:
\begin{equation}\label{eq:resultOfInterference4}
S = 4E_0^2{\cos ^2}\left[ {\frac{1}{2}({\omega _0}\tau  + \Delta \varphi )} \right],
\end{equation}
where  $\Delta \varphi  = \varphi _2^0 - \varphi _1^0$. In the general case, the phase difference $\Delta \varphi$  is a random function of time, and for incoherent beams it fluctuates very quickly, so that the interference cannot be observed. In the case of independent quasi-monochromatic laser beams, $\Delta \varphi$  is still a function of time, but its variations occur quite slowly, so that the interference can be easily observed.

Note that if the two independent laser beams with fields, as in Eq.~\eqref{eq:fieldsE1E2}, are brought into the same input port of an unbalanced interferometer, the result of the interference will be determined by the following relation:
\begin{equation}\label{eq:resultOfInterference2}
S = 4E_0^2{\cos ^2}\left( {\frac{{{\omega _0}\Delta T}}{2}} \right){\cos ^2}\left[ {\frac{1}{2}({\omega _0}\tau  + \Delta \varphi )} \right].
\end{equation}
An important difference between the interference of independent coherent laser beams (Eq.~\eqref{eq:resultOfInterference2}) from the interference of a monochromatic wave with itself (Eq.~\eqref{eq:resultOfInterference1}) is that the former depends on the phase difference  $\Delta\varphi$, as well as on the additional phase change $\omega_0\tau$  associated with the "prehistory" of the interfering beams.

Let us now consider the interference of pulses from a semiconductor laser when the pump current does not fall below threshold, i.e., when the laser operates in a continuous mode, and the modulation current “cuts out” pulses from the continuous laser beam (we neglect the effect of chirp). We will further assume that the sequence of laser pulses obtained in this way is fed into the unbalanced interferometer with the time delay equal to the pulse repetition period ($\Delta T=1/f_p$; $f_p$ is the pulse repetition frequency). With such an interferometer, we will observe the interference of the two neighboring pulses. Obviously, this case is similar to the interference of a monochromatic wave with itself with the only difference that the envelopes of the interfering components now depend on time, and instead of Eq.~\eqref{eq:fieldsInTermsOfE0} we should write:
\begin{equation}\label{eq:fieldsInTermsOfQ}
\begin{split}
{E_1}(t) &= \sqrt {{Q_1}(t)} \exp \left( {i{\varphi _0} + i{\omega _0}t} \right),\\
{E_2}(t) &= \sqrt {{Q_2}(t)} \exp \left[ {i{\varphi _0} + i{\omega _0}(t + \Delta T)} \right],
\end{split}
\end{equation}
where $Q_1$ and $Q_2$ are field intensities. Here, we will assume that optical pulses have a Gaussian shape:
\begin{equation}\label{eq:Q1Q2}
{Q_1}(t) = {Q_0}{e^{ - \frac{{{t^2}}}{{2{w^2}}}}},\,\,\,\,\,
{Q_2}(t) = {Q_0}{e^{ - \frac{{{{(t - \Delta t)}^2}}}{{2{w^2}}}}},
\end{equation}
where $w$  is the width of the pulse and $\Delta t$  takes into account the inaccuracy of the pulse overlap. In the general case, $\Delta t$  can be associated with a difference between the pulse repetition period and the time delay in the interferometer, as well as with the time jitter of laser pulses. Below, we will assume that $\Delta t$  is associated only with jitter, whereas the time delay $\Delta T$ in the long arm of the interferometer ideally matches the pulse repetition period.

The result of the interference of laser pulses with fields from Eq.~\eqref{eq:fieldsInTermsOfQ} is
\begin{equation}\label{eq:resultOfInterference3}
S = {Q_1} + {Q_2} + 2\sqrt {{Q_1}{Q_2}} \cos ({\omega _0}\Delta T).
\end{equation}
In the case of the pulse interference, it is useful to determine the integral signal $\tilde{S}$ corresponding to the "area" under the interference pulse normalized to the "area" of the signal passing through one of the interferometer arms:
\begin{equation}\label{eq:normalizedSignal}
\tilde S = \frac{{\int\limits_{ - {{\Delta T}/2}}^{{{\Delta T}/2}} {S(t)dt} }}{{\int\limits_{ - \Delta {T/2}}^{{{\Delta T}/2}} {{Q_1}(t)dt} }},
\end{equation}
whence using Eqs.~\eqref{eq:Q1Q2} and \eqref{eq:resultOfInterference3} and extending integration limits up to $\pm\infty$  (this can be done if the width  $w$ is significantly smaller than the pulse repetition period $\Delta T$) we will have:
\begin{equation}\label{eq:resultOfInterference6}
\tilde S = 2\left[ {1 + \eta \cos ({\omega _0}\Delta T)} \right],
\end{equation}
where visibility  $\eta=\exp[-{{\Delta t^2}/(8{w^2})}]$ depends on the ratio between jitter $\Delta t$  and the pulse width  $w$. It is important to note here that visibility of the interference in this extreme case does not depend on the spectral composition of light, but depends only on  jitter, i.e. on the quality of the electrical pattern that sets the sequence of laser pulses. However, if the time jitter is small, $\Delta t\ll w$, one may assume that  $\eta  \approx 1$, which yields  $\tilde S = 4{\cos ^2}({{{\omega _0}\Delta T}/2})$, and the result of the pulse interference is determined by the phase evolution ${\omega _0}\Delta T$  and \textit{does not depend} on jitter.

Now let us consider the above scheme of laser pulse interference ($\Delta T = {{1 }/{{f_p}}}$), but now with the semiconductor laser operating under the gain switching. If the coherence of radiation in the cavity is destroyed during the time while the laser is under the threshold, then we may assume that the pulses meeting at the interferometer’s output originate from independent sources. The result of the interference, therefore, should be similar to that obtained in Eq.~\eqref{eq:resultOfInterference4}. In fact, by analogy with Eq.~\eqref{eq:fieldsE1E2}, let us write the fields in the interfering pulses:
\begin{equation}\label{eq:fieldsInTermsOfQ1Q2}
\begin{split}
{E_1}(t)& = \sqrt {{Q_1}(t)} \exp \left( {i\varphi _1^0 + i{\omega _0}t} \right),\\
{E_2}(t)& = \sqrt {{Q_2}(t)} \exp \left[ {i\varphi _2^0 + i{\omega _0}(t + \Delta t)} \right],
\end{split}
\end{equation}
where the phase change  ${\omega _0}\tau$ from Eq.~\eqref{eq:fieldsE1E2} is replaced here by  ${\omega _0}\Delta t$ since the "prehistory" is related in this case with jitter, due to which different pairs of pulses do not always arrive at the second beam splitter simultaneously. Moreover, we will assume that $Q_1$  and $Q_2$ in Eq.~\eqref{eq:fieldsInTermsOfQ1Q2} are again defined by Eq.~\eqref{eq:Q1Q2}.

Let us first consider the case when there is no jitter. The integral signal $\tilde{S}$  will then have a simple form:
\begin{equation}\label{eq:resultOfInterference5}
\tilde S = 4{\cos ^2}\frac{{\Delta \varphi }}{2}.
\end{equation}
As can be seen from a comparison of Eqs.~\eqref{eq:resultOfInterference5} and \eqref{eq:resultOfInterference6}, the result of the interference of independent laser pulses does not include the carrier frequency of the electromagnetic field. This means that if we take a laser of another wavelength and use it to prepare a similar pair of pulses with the same initial phases, then the result of the interference of this new pair of pulses will be the same as in the previous case. This result differs significantly from the interference of a continuous monochromatic wave in an unbalanced interferometer, for which the shift of the carrier frequency leads to the change in the result of the interference.

Now let us consider the interference of independent pulses in the presence of jitter. In this case, the result of the interference of fields from Eq.~\eqref{eq:fieldsInTermsOfQ1Q2} will have the following form:
\begin{equation}\label{eq:resultOfInterference7}
\tilde S = 2\left[ {1 + \eta \cos ({\omega _0}\Delta t + \Delta \varphi )} \right].
\end{equation}
If  $\Delta t \ll w$, we may put  $\eta\approx 1$, which yields  $\tilde S = 4{\cos ^2}\left[ {{{({\omega _0}\Delta t + \Delta \varphi )}/2}} \right]$, whence it is clear that the result of the interference is strongly dependent on jitter since the latter is included in the cosine argument. Moreover, for a fixed  $\Delta\varphi$, the interference of pulses in this case will, in essence, be determined by jitter. Indeed, for frequencies commonly used in telecommunications, ${{{\omega _0}}/(2\pi)} \sim {10^{14}}$   Hz, even highly stable frequency oscillators with jitter of hundreds of femtoseconds, $\Delta t\sim{10^{ - 13}}$, will not allow obtaining the stable interference (at fixed  $\Delta\varphi$), because  ${\omega _0}\Delta t \gg 1$. It should be noted here that the time jitter of laser pulses from a gain-switched laser is generally much larger than the intrinsic jitter of the driving electrical signal. This fact is related to a delay between the leading edge of the pump current pulse and the onset of lasing (the so-called turn-on delay \cite{Konnerth64}), which depends on the amplitude (peak-to-peak value  $I_p$) of the modulation current. Due to fluctuations in  $I_p$, the turn-on delay will also fluctuate, which will lead to an additional contribution to jitter. 

Comparing Eqs.~\eqref{eq:resultOfInterference7} and \eqref{eq:resultOfInterference6}, we may write the following condition (in assumption that the influence of jitter on visibility can be neglected):
\begin{equation}\label{eq:transition}
4{\cos ^2}\left( {\frac{{{\omega _0}\Delta t + \Delta \varphi }}{2}} \right) \to 4{\cos ^2}\left( {\frac{{{\omega _0}\Delta T}}{2}} \right),
\end{equation}
which determines the transition of the laser from the gain switching mode to continuous generation. It is important to note that the left-hand side of Eq.~\eqref{eq:transition} is valid when the bias current $I_b$ is significantly below threshold current $I_{th}$, whereas the right-hand side is valid when $I_b$  is much higher than  $I_{th}$. Obviously, the transition given by Eq.~\eqref{eq:transition} is equivalent to the following conditions: $\Delta \varphi  \to {\omega _0}\Delta T$  and  $\Delta t \to 0$, which are satisfied if we assume that $\Delta\varphi$  and $\Delta t$ are random variables with Gaussian distributions, whose average values are $\omega_0\Delta t$  and 0, respectively, and whose standard deviations ${\sigma _\varphi } \equiv {\sigma _\varphi }({I_b})$ and  ${\sigma _{\Delta t}} \equiv {\sigma _{\Delta t}}({I_b})$ are decreasing functions of the bias current, and they tend to zero when $I_b$ is much higher than  $I_{th}$.

In a QRNG based on the interference of laser pulses, the phase difference $\Delta\varphi$  plays the role of a quantum entropy source since its random values are determined by quantum nature of spontaneous emission. In contrast, the jitter-related phase change  $\omega_0\Delta t$ should be treated as classical noise. For typical jitter values ($10^{-12}$~s), the phase evolution ${\omega _0}\Delta t$  can take on very large values, significantly exceeding  $\Delta\varphi$; therefore, at first glance, it seems that randomness obtained from the interference of laser pulses cannot be considered quantum. This could be true if, with a large range of ${\omega _0}\Delta t$  values, the spread of  $\Delta\varphi$ values was small enough. However, the influence of jitter on the phase difference between laser pulses can be ignored at small values of $\sigma_\varphi$  (small spread of $\Delta\varphi$  values). Indeed, in this case, adjacent laser pulses cannot be considered as pulses from two independent sources since they are now phase correlated. This means that such pulses should be considered "cut out" from a continuous quasi-monochromatic wave (albeit with a short coherence time), for which the effect of jitter on the phase evolution is absent.

The above reasoning shows that substantiation of quantum nature of laser pulse interference is a "thin place" of QRNG implementation. Indeed, continuing the above reasoning, one can come to the conclusion that for a sufficiently large value of $\sigma_\varphi$  (at least for $\sigma_\varphi>\pi$) interfering laser pulses can be considered independent and one should take into account the jitter-related phase change ${\omega _0}\Delta t$. Inasmuch as classical contribution from ${\omega _0}\Delta t$  can significantly exceed quantum contribution from  $\Delta\varphi$, a reasonable question arises: does the QRNG cease to be quantum in this case? In fact, it is easy to see that QRNG still should be considered quantum. Indeed, inasmuch as the sum ${\omega _0}\Delta t + \Delta \varphi$  is in the argument of the cosine, an adversary who knows all the $\Delta t$  values in advance or may even control them will not be able to say anything definite about the result of the interference (beyond what he a priori knows about the probability density of the interference signal), if  $\Delta\varphi$  exhibits a large spread of values. In other words, even in the presence of a classical contribution from ${\omega _0}\Delta t$  the result of the interference from Eq.~\eqref{eq:resultOfInterference7} remains nondeterministic, unpredictable and uncontrollable, i.e. satisfies the requirements for the quantum noise \cite{Shakhovoy2020}. This means that when considering the interference of laser pulses with random phases in the context of a QRNG, the phase change associated with jitter can be excluded from consideration.

Finally, note that the result of interference is also influenced by the spread in the intensities of laser pulses associated with fluctuations of the pump current. We will assume below that such fluctuations are sufficiently small, so that we can neglect the change in the shape of the laser pulse and assume that only its "area" changes slightly from pulse to pulse. Thus, the fields in interfering pulses can be written in the following form:
\begin{equation}\label{eq:fields2}
\begin{split}
{E_1}(t)& = \sqrt {{s_1}{Q_1}(t)} \exp \left( {i\varphi _1^0 + i{\omega _0}t} \right),\\
{E_2}(t)& = \sqrt {{s_2}{Q_2}(t)} \exp \left( {i\varphi _2^0 + i{\omega _0}t} \right),
\end{split}
\end{equation}
where we have introduced two independent random variables, $s_1$  and $s_2$, associated with the spread of intensities of interfering pulses. For simplicity, we may assume that both  $s_1$  and $s_2$ exhibit Gaussian distribution with the mean value ${\bar s_1} = {\bar s_2} = 1$  and a standard deviation of  $\sigma_s$. The integral interference signal $\tilde{S}$  for the fields from Eq.~\eqref{eq:fields2} will then have the form:
\begin{equation}\label{eq:resultOfInterference8}
\tilde S = {s_1} + {s_2} + 2\eta \sqrt {{s_1}{s_2}} \cos (\Delta \varphi).
\end{equation}

\subsection{Probability density function and\\ the quantum reduction factor}
In \cite{Shakhovoy2021}, the influence of chirp, jitter, and relaxation oscillations on the probability density of the integral interference signal $\tilde{S}$ has been considered in assumption that $\Delta\varphi$  values are distributed uniformly in the range $[0,\pi)$. In particular, the chirp has been shown to enhance the influence of jitter on visibility $\eta$, so that the condition  $\Delta t\ll w$ does not always allow neglecting fluctuations of  $\eta$. Nevertheless, it was shown that the "chirp + jitter" effect could be significantly decreased for relatively short laser pulses by cutting off the high-frequency part of the spectrum. As for sufficiently long laser pulses, the effect of the chirp on the interference can be obviously neglected (even without spectral filtering) if the "area" under the main relaxation peak is much smaller than the “area” under the rest of the pulse. With this in mind, we will assume below that we deal with quite long laser pulses or use spectral filtering, so that the chirp has no significant effect on the interference. We will thus neglect jitter-related fluctuations of the visibility. In this case, the probability distribution function (PDF) ${f_{\tilde S}}$  of the integral signal $\tilde{S}$  can be defined as a derivative of a corresponding cumulative distribution function. The latter represents an integral of the PDF defining joint fluctuations of  $\Delta\varphi$,  $s_1$,  $s_2$, and  $\zeta$, where $\zeta$ is the Gaussian classical detector noise, which should be added to the integral signal:  $\tilde S \to \tilde S + \zeta $. Generally, an analytical expression for $f_{\tilde S}$  cannot be found; therefore, the analysis of various contributions to $f_{\tilde S}$  is performed numerically with Monte-Carlo simulations.

\begin{figure}[t]
	\includegraphics[width=\columnwidth]{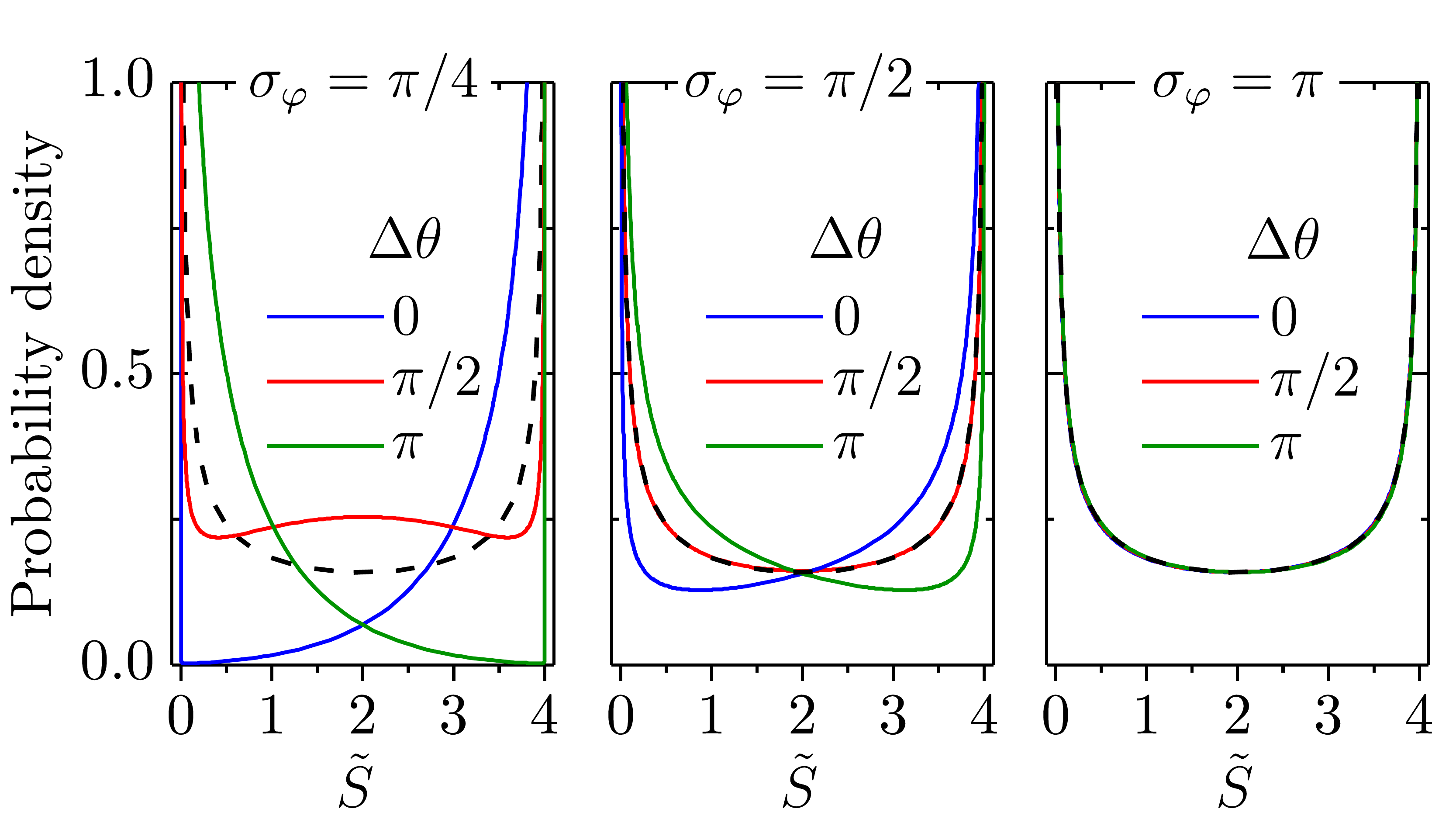}
	\caption{\label{fig:PDFsIdeal} Probability density functions $\tilde{f}_{\tilde{S}}^Q$ defined by Eq.~\eqref{eq:quantumPDFWithGaussian3} at different values of $\sigma_{\varphi}$ and $\Delta\theta$.}
\end{figure}

It was shown in \cite{Shakhovoy2020} that one can extract quantum noise from the interference signal by introducing the parameter called the quantum reduction factor (QRF)  $\Gamma$. This parameter shows how much the raw random sequence should be reduced (or compressed) by the randomness extractor in order to filter out possible hidden correlations associated with classical noise. The idea of the QRF is based on a comparison of the experimentally observed PDF $f_{\tilde S}$  with ideal (or quantum) PDF  $f_{\tilde S}^Q$. The latter can be found from Eq.~\eqref{eq:resultOfInterference8} in assumption that the only random variable is $\Delta\varphi$  (whereas $s_1$  and  $s_2$ are fixed to their average values), and
\begin{equation}\label{eq:uniformPDF}
{f_{\Delta \varphi }} = \left\{ {\begin{array}{*{20}{c}}
	{1/\pi,\,\Delta \varphi  \in [0,\pi )}\\
	{0,\,\,\,\Delta \varphi  \notin [0,\pi )}
	\end{array}} \right..
\end{equation}
One can easily show that
\begin{equation}\label{eq:quantumPDF}
f_{\tilde S}^Q(y) = {\left[ {\pi \sqrt {(y - {{\tilde S}_{min}})({{\tilde S}_{max}} - y)} } \right]^{ - 1}},
\end{equation}
where
\begin{equation}
\begin{split}
{{\tilde S}_{min}}& = {s_1} + {s_2} - 2\eta \sqrt {{s_1}{s_2}} ,\\
{{\tilde S}_{max}}& = {s_1} + {s_2} + 2\eta \sqrt {{s_1}{s_2}} .
\end{split}
\end{equation}
The function given by Eq.~\eqref{eq:quantumPDF} with  ${s_1} = {s_2} = 1$ and $\eta=1$  is shown in Fig.~\ref{fig:PDFsIdeal} by black dashed lines; one can see that $f_{\tilde S}^Q$  is U-shaped, and tends to infinity at ${\tilde S_{min}}$  and ${\tilde S_{max}}$  (in the case under consideration, ${\tilde S_{min}}=0$  and ${\tilde S_{min}}=4$).

It was shown that the definition of $\Gamma$  depends on the method of digitizing the random signal \cite{Shakhovoy2020}. When digitizing with a comparator, one can define   as follows:
\begin{equation}\label{eq:quantumReductionFactor}
\Gamma  = \frac{1}{{2 - {H_\infty }}},
\end{equation}
where the min-entropy is defined as
\begin{equation}\label{eq:minEntropy}
{H_\infty } =  - {\log _2}\left( {\int\limits_{{{\tilde S}_{min}}}^{{{\tilde S}_{th}}} {{f_{\tilde S}}(y)dy} } \right),
\end{equation}
and $\tilde{S}_{th}$  is a threshold value, which should be chosen such that the area under the  $f_{\tilde{S}}(y)$ curve to the left and to the right of $\tilde{S}_{th}$  was  $1/2$. (With such a definition, we may also treat $\tilde{S}_{th}$ as a mean value of the distribution.) Note that the quantum PDF $f_{\tilde{S}}^Q$  is implicitly contained in Eq.~\eqref{eq:minEntropy} via the lower limit of the integral. If we substitute $f_{\tilde{S}}^Q$  instead of $f_{\tilde{S}}$  into Eq.~\eqref{eq:minEntropy}, the min-entropy will have the sense of ideal or quantum min-entropy, which we denote as  $H_\infty^Q$; obviously,  ${H_\infty ^Q = 1}$.

It is important to keep in mind that the definition of the QRF given by Eqs.~\eqref{eq:quantumReductionFactor} and \eqref{eq:minEntropy} is directly related to the assumption that  ${f_{\Delta \varphi }}$ is uniform. In the general case, however, ${f_{\Delta \varphi }}$  is Gaussian:
\begin{equation}\label{eq:gaussianPDF}
{f_{\Delta \varphi }}(x) = \frac{1}{{{\sigma _\varphi }\sqrt {2\pi } }}\exp \left( { - \frac{{{{(x - \Delta \theta )}^2}}}{{2\sigma _\varphi ^2}}} \right),
\end{equation}
where  $\Delta \theta  = {\omega _0}\Delta T$, and the value of the random variable $\Delta\varphi$  we have denoted as  $x$. To find the quantum PDF $f_{\tilde{S}}^Q$   in this case, we use a well-known theorem applicable to monotonic functions, according to which the probability density ${f_Y}(y)$  of a random variable $Y = g(X)$  is given by the following formula: 
\begin{equation}\label{eq:formulaForProbabilityDensity}
{f_Y}(y) = {f_X}\left( {{g^{ - 1}}(y)} \right)\left| {\frac{d}{{dy}}\left( {{g^{ - 1}}(y)} \right)} \right|,
\end{equation}
where  $y$ is the value of a random variable  $Y$, $f_X$  is a PDF of a random variable  $X$, and  ${g^{ - 1}}(y)$ is the inverse function. To apply this theorem to a random function $\tilde S(\Delta \varphi )$  defined by Eq.~\eqref{eq:resultOfInterference8}, we have to divide the domain of $\tilde S(\Delta \varphi )$  into intervals, where it is piecewise monotonic. Obviously, $\tilde S(\Delta \varphi )$   is monotonically decreasing in the intervals $\Delta \varphi  \in I_ \downarrow ^m \equiv [2m\pi ,(2m + 1)\pi )$  ($m$  is integer), whereas it is monotonically increasing in the intervals $\Delta \varphi  \in I_ \uparrow ^m \equiv [(2m - 1)\pi ,2m\pi )$  (we introduced here the notation    $I_ \downarrow ^m$  and  $I_ \uparrow ^m$   for the intervals of monotonicity). In the intervals of monotonicity, there exist inverse functions, which we will denote as $\tilde S_{I_ \downarrow ^m}^{ - 1} \equiv \tilde S_ \downarrow ^{ - 1}(y)$ and $\tilde S_{I_ \uparrow ^m}^{ - 1} \equiv \tilde S_ \uparrow ^{ - 1}(y)$. It is easy to see from Eq.~\eqref{eq:resultOfInterference8} that
\begin{equation}\label{eq:inverseFunction}
\tilde S_{ \downarrow , \uparrow }^{ - 1}(y) =  \pm \arccos \left( {\frac{{2y - {{\tilde S}_{max}} - {{\tilde S}_{min}}}}{{{{\tilde S}_{max}} - {{\tilde S}_{min}}}}} \right) + 2\pi m,
\end{equation}
where the "plus" sign refers to  $\tilde S_ \downarrow ^{ - 1}$, and the "minus" sign refers to  $\tilde S_ \uparrow ^{ - 1}$ (the value of the random variable $\tilde{S}$  we have denoted as  $y$). According to Eq.~\eqref{eq:formulaForProbabilityDensity}, the PDF of a piecewise monotonic function defined in such a way can be written in the following form:
\begin{equation}\label{eq:quantumPDFWithGaussian}
\tilde f_{\tilde S}^Q = \sum\limits_{i = I_ \downarrow ^m,I_ \uparrow ^m} {{f_{\Delta \varphi }}(\tilde S_i^{ - 1})|(\tilde S_i^{ - 1})'|} \,,
\end{equation}
where
\begin{equation}\label{eq:derivative}
\begin{split}
(\tilde S_{ \downarrow , \uparrow }^{ - 1})' &\equiv \frac{d}{{dy}}\tilde S_{ \downarrow , \uparrow }^{ - 1}(y)\\
& =  \mp {\left[ {\sqrt {(y - {{\tilde S}_{min}})({{\tilde S}_{max}} - y)} } \right]^{ - 1}},
\end{split}
\end{equation}
and where the "minus" sign refers to the derivative of  $\tilde S_ \downarrow ^{ - 1}$, whereas the "plus" sign refers to the derivative of  $\tilde S_ \uparrow ^{ - 1}$. Substituting Eq.~\eqref{eq:derivative} into Eq.~\eqref{eq:quantumPDFWithGaussian} and using Eqs.~\eqref{eq:gaussianPDF} and \eqref{eq:inverseFunction} we will obtain:
\begin{equation}\label{eq:quantumPDFWithGaussian2}
\begin{split}
&\tilde f_{\tilde S}^Q(y) = {\left[ {{\sigma _\varphi }\sqrt {2\pi (y - {{\tilde S}_{min}})({{\tilde S}_{max}} - y)} } \right]^{ - 1}}\\
&\times \sum\limits_{p =  \pm 1} {\sum\limits_{m =  - \infty }^{ + \infty } {\exp \left( { - \frac{1}{{2\sigma _\varphi ^2}}{{\left[ {p{a_y} + 2\pi m - \Delta \theta } \right]}^2}} \right)} } ,
\end{split}
\end{equation}
where we used the shorthand notation
\begin{equation}
{a_y} = \arccos \left( {\frac{{2y - {{\tilde S}_{max}} - {{\tilde S}_{min}}}}{{{{\tilde S}_{max}} - {{\tilde S}_{min}}}}} \right).
\end{equation}
The sum in Eq.~\eqref{eq:quantumPDFWithGaussian2} converges to
\begin{equation}\label{eq:quantumPDFWithGaussian3}
\tilde f_{\tilde S}^Q(y) = \sum\limits_{p =  \pm 1} {\frac{{J\left( {\frac{{p{a_y}}}{2} - \frac{{\Delta \theta }}{2},{e^{ - {{\sigma _\varphi ^2}/ 2}}}} \right)}}{{2\pi \sqrt {(y - {{\tilde S}_{min}})({{\tilde S}_{max}} - y)} }}},
\end{equation}
where $J(u,q)$  is the Jacobi theta function:
\begin{equation}\label{eq:jacobiThetaFunction}
J(u,q) = 1 + 2\sum\limits_{j = 1}^\infty  {{q^{{j^2}}}\cos (2ju)} .
\end{equation}

An important difference between PDFs given by Eqs.~\eqref{eq:quantumPDF} and \eqref{eq:quantumPDFWithGaussian3} is that the latter depends on  $\Delta \theta$. The form of the $\tilde{f}_{\tilde S}^Q$  function at various values of $\Delta \theta$  for the cases ${\sigma _\varphi } = {\pi/4},\, {\pi/2},\, \pi$  is shown in Fig.~\ref{fig:PDFsIdeal}, where it is assumed that ${s_1} = {s_2} = 1$  and  $\eta  = 1$. When  ${\sigma _\varphi } = {\pi/4}$, the function $\tilde f_{\tilde S}^Q$  substantially differs from  $f_{\tilde S}^Q$ at any value of $\Delta\theta$  (recall that the  $f_{\tilde S}^Q$ function given by Eq.~\eqref{eq:quantumPDF} is shown by the black dashed line). The values $\Delta \theta  = 0$  and $\Delta \theta  = \pi$ yield in the substantial shift of the PDF into the region of constructive (blue line) and destructive (green line) interference, respectively, whereas at $\Delta \theta  = {\pi/2}$ the PDF exhibits a maximum at  $\tilde S = 2$. When  ${\sigma _\varphi } = {\pi/2}$, the shift is still clearly visible at $\Delta \theta  = 0$  and  $\Delta \theta  = \pi$, whereas $\tilde f_{\tilde S}^Q$  and $f_{\tilde S}^Q$  become almost indistinguishable at  $\Delta \theta  = {\pi/2}$. Finally, when  ${\sigma _\varphi } = \pi$, the $\tilde f_{\tilde S}^Q$  function is weakly dependent on $\Delta \theta$  and is almost indistinguishable from $f_{\tilde S}^Q$  at any value of  $\Delta \theta$.

\begin{figure}[t]
	\includegraphics[width=0.8\columnwidth]{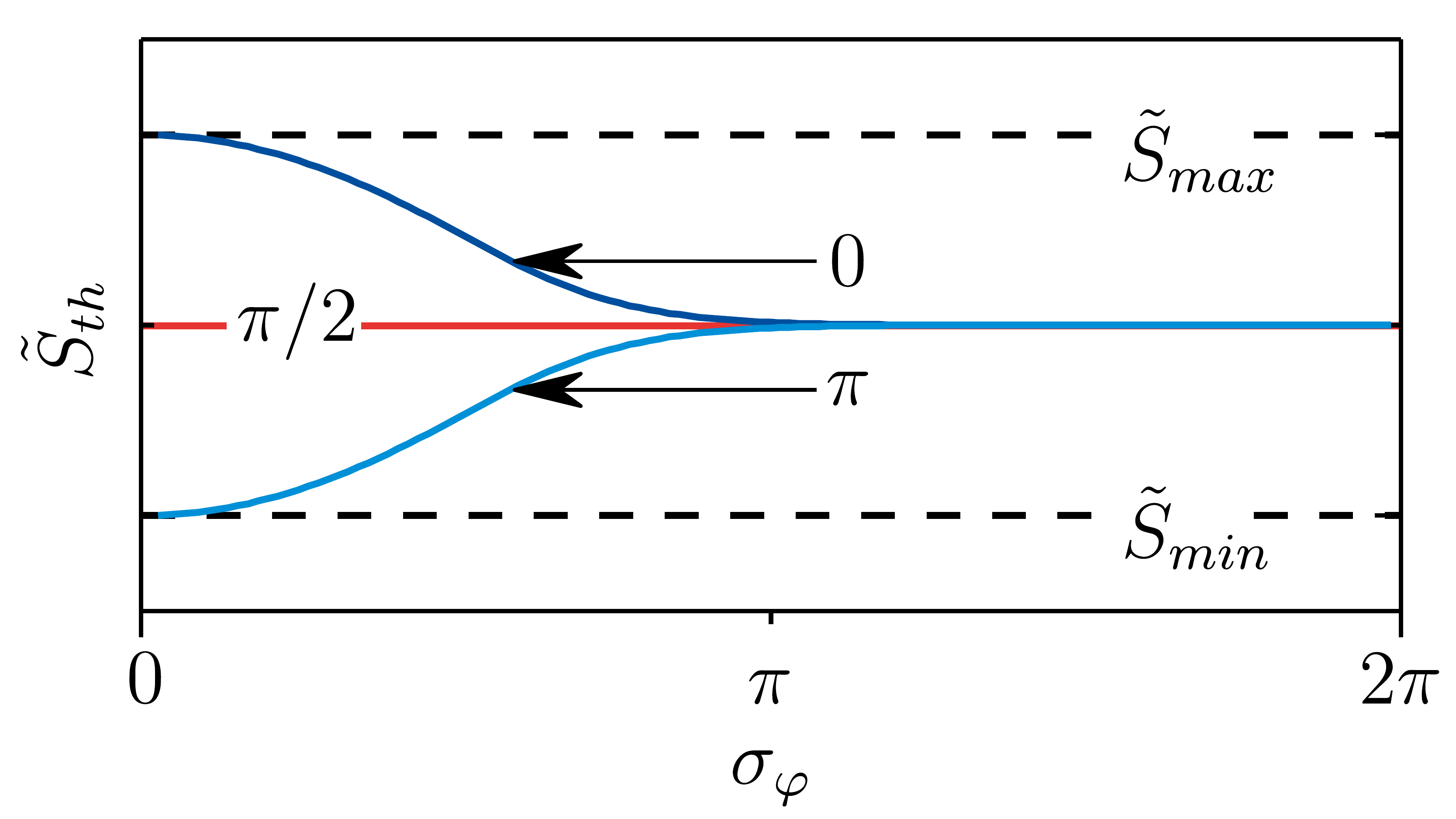}
	\caption{\label{fig:Sth} Calculated dependences ${\tilde S_{th}}({\sigma _\varphi })$  for the three values of  $\Delta\theta$: 0, $\pi/2$, and $\pi$.}
\end{figure}

Let us return to the definition of the QRF. At first glance, we are not prohibited to use of formulas \eqref{eq:quantumReductionFactor} and \eqref{eq:minEntropy} in the case when $\sigma_\varphi$ is quite small. The only difference here would be the dependence of the threshold value  ${\tilde S_{th}}$ in Eq.~\eqref{eq:minEntropy} on both $\sigma_\varphi$  and  $\Delta\theta$. To clarify this, let us consider the calculated dependences ${\tilde S_{th}}({\sigma _\varphi })$  for the three values of  $\Delta\theta$ shown in Fig.~\ref{fig:Sth}. One can see from the figure that ${\tilde S_{th}}$  is almost independent on both $\sigma_\varphi$  and $\Delta\theta$  when  $\sigma_\varphi>\pi$; in addition, the threshold value remains equal to ${{({{\tilde S}_{max}} + {{\tilde S}_{min}})}/2}$ and does not depend on  $\sigma_\varphi$, when  $\Delta \theta  = {\pi/2}$. However, ${\tilde S_{th}}$  differs significantly from ${{({{\tilde S}_{max}} + {{\tilde S}_{min}})}/2}$  at smaller values of  $\sigma_\varphi$, if $\Delta\theta$ is far from  $\pi/2$. Thus, ${\tilde S_{th}} \to {\tilde S_{max}}$  at $\Delta \theta  = 0$  and ${\tilde S_{th}} \to {\tilde S_{min}}$  at $\Delta \theta  = \pi$ when $\sigma_\varphi$  approaches zero. It seems that such a dependence of ${\tilde S_{th}}$ on $\sigma_\varphi$  and  $\Delta\theta$ should not affect the quantum noise extraction method itself since ${\tilde S_{th}}$  is explicitly included in the definition of  $\Gamma$. Nevertheless, one can show that the use of Eqs.~\eqref{eq:quantumReductionFactor} and \eqref{eq:minEntropy} is hardly possible when  ${\sigma _\varphi }<\pi$.

The main reason for this statement is that $\tilde f_{\tilde S}^Q$  function averaged over $\Delta \theta$  becomes equal to  $f_{\tilde S}^Q$:
\begin{equation}\label{eq:averaging}
\frac{1}{{2\pi }}\int\limits_0^{2\pi } {\tilde f_{\tilde S}^Qd(\Delta \theta )}  = f_{\tilde S}^Q.
\end{equation}
It follows from Eq.~\eqref{eq:averaging} that an adversary may "mimic" the large-$\sigma_\varphi$  PDF via controlling  $\Delta \theta$, such that the user employing  ${f_{\tilde S}}$ to monitor the contribution of classical noise will believe that he is still working with a quantum entropy source. In fact, the phase  $\Delta\theta$ in the long arm of the interferometer should be considered as a "classical" parameter, which can be predicted or even controlled (at least in principle) by a third party. Indeed, since  $\Delta\theta$ is related to the optical path difference of the interferometer arms, it is enough for an adversary to control the temperature near the device to get insight into the value of the signal  $\tilde{S}$. Thus, using the dependence of ${\tilde S_{th}}$  on $\Delta\theta$, an adversary may predict (with a probability different from  $1/2$) each bit of the random sequence digitized by a comparator when  ${\sigma _\varphi }<\pi$. 

In contrast, it can be shown that the PDF of the phase difference $\Delta\varphi$  may be assumed uniform with high accuracy, when  ${\sigma _\varphi } > 2\pi$; in other words, we may neglect the difference between $\tilde f_{\tilde S}^Q$  and $f_{\tilde S}^Q$  in this case. Indeed, inasmuch as $\Delta\varphi$  is in the argument of the cosine in Eq.~\eqref{eq:resultOfInterference8}, the Gaussian PDF ${f_{\Delta \varphi }}$  from Eq.~\eqref{eq:gaussianPDF} may be substituted for $x\in [0,\pi)$ by the following one (see Appendinx in \cite{Shakhovoy2020}):
\begin{equation}\label{eq:PDFInTermsOfJ}
\tilde f_{\Delta \varphi }(x) = \frac{1}{{2\pi }}\sum\limits_{p =  \pm 1} J\left( {\frac{x}{2} + \frac{{p\Delta \theta }}{2},{e^{ - {{\sigma _\varphi ^2}/2}}}} \right)
\end{equation}
and should be put to 0 when ${x \notin [0,\pi )}$. Here $J(u,q)$ is again the Jacobi theta function defined by Eq.~\eqref{eq:jacobiThetaFunction}. Inasmuch as $q<1$  in this case, the series in Eq.~\eqref{eq:PDFInTermsOfJ} rapidly converges, so the value of the theta function may be estimated with just the two first terms:  $J(u,q) \approx 1 + 2q\,\cos (2u)$, whence one can see that $J(u,q)$  deviates from 1 by a value  $2q\sim{10^{ - 8}}$ at  ${\sigma _\varphi } = 2\pi$.

Thus, we may conclude that fluctuations of the phase difference $\Delta\varphi$  in Eq.~\eqref{eq:resultOfInterference8} can be considered quantum when  ${\sigma _\varphi } > 2\pi$, whereas they become highly sensitive to variations of $\Delta\theta$  when ${\sigma _\varphi } < \pi$  and cannot be used as a quantum entropy source. But what about the range from $\pi$  to  $2\pi$? We may just look at Fig.~\ref{fig:PDFsIdeal} and Fig.~\ref{fig:Sth} and repeat again that $\tilde f_{\tilde S}^Q$  is almost indistinguishable from  $f_{\tilde S}^Q$, when $\sigma_{\varphi}$ is in this range. At first sight, one may just soften the restriction imposed on the values of $\sigma_\varphi$  and assume for simplicity that everything still works and the formulas \eqref{eq:quantumPDF}, \eqref{eq:quantumReductionFactor} and \eqref{eq:minEntropy} are valid when  ${\sigma _\varphi } > \pi$. Nevertheless, it would be useful to have a quantitative estimate for the difference between $\tilde f_{\tilde S}^Q$  and  $f_{\tilde S}^Q$ when $\sigma_\varphi$  is in the range from  $\pi$ to  $2\pi$, and, if necessary, to modify the QRF in order to take into account the possible influence of an adversary.

In the information theory, the difference between the two distributions, $P_X$  and  $P_Y$, is generally measured in terms of a statistical distance  $d$, which for the countable set $\Omega$  of elementary events can be defined as follows \cite{Nisan99}:
\begin{equation}\label{eq:statisticalDistance}
d = \frac{1}{2}\sum\limits_{a \in \Omega } {\left| {{P_X}(a) - {P_Y}(a)} \right|},
\end{equation}
where $P_{X(Y)}(a)$  is a probability that the random variable $X(Y)$  takes a value  $a$. By analogy with Eq.~\eqref{eq:statisticalDistance}, the statistical distance between the PDFs  $\tilde f_{\tilde S}^Q$ and $f_{\tilde S}^Q$ can be written in the form of the following integral:
\begin{equation}\label{eq:statisticalDistanceForPDFs}
d = \frac{1}{2}\int\limits_{{{\tilde S}_{min}}}^{{{\tilde S}_{max}}} {\left| {\tilde f_{\tilde S}^Q(y) - f_{\tilde S}^Q(y)} \right|dy} ,
\end{equation}
where $\tilde f_{\tilde S}^Q$ and $f_{\tilde S}^Q$ are defined by Eqs.~\eqref{eq:quantumPDFWithGaussian3} and \eqref{eq:quantumPDF}, respectively. It is not very convenient, however, to calculate numerically the integral in Eq.~\eqref{eq:statisticalDistanceForPDFs} since $\tilde f_{\tilde S}^Q$ and $f_{\tilde S}^Q$   have singularities at ${\tilde S_{min}}$  and  ${\tilde S_{max}}$; therefore, it is reasonable to define $d$ as a statistical distance between ${\tilde f_{\Delta \varphi }}$   and  ${f_{\Delta \varphi }}$, namely:
\begin{equation}\label{eq:statisticalDistanceForPDFs2}
d = \frac{1}{2}\int\limits_0^\pi  {\left| {{{\tilde f}_{\Delta \varphi }}(x) - {\pi ^{ - 1}}} \right|dx} ,
\end{equation}
where ${\tilde f_{\Delta \varphi }}$  is defined by Eq.~\eqref{eq:PDFInTermsOfJ}. One can easily check (e.g., numerically) that both definitions, Eq.~\eqref{eq:statisticalDistanceForPDFs} and Eq.~\eqref{eq:statisticalDistanceForPDFs2}, are equivalent.

\begin{figure}[t]
	\includegraphics[width=\columnwidth]{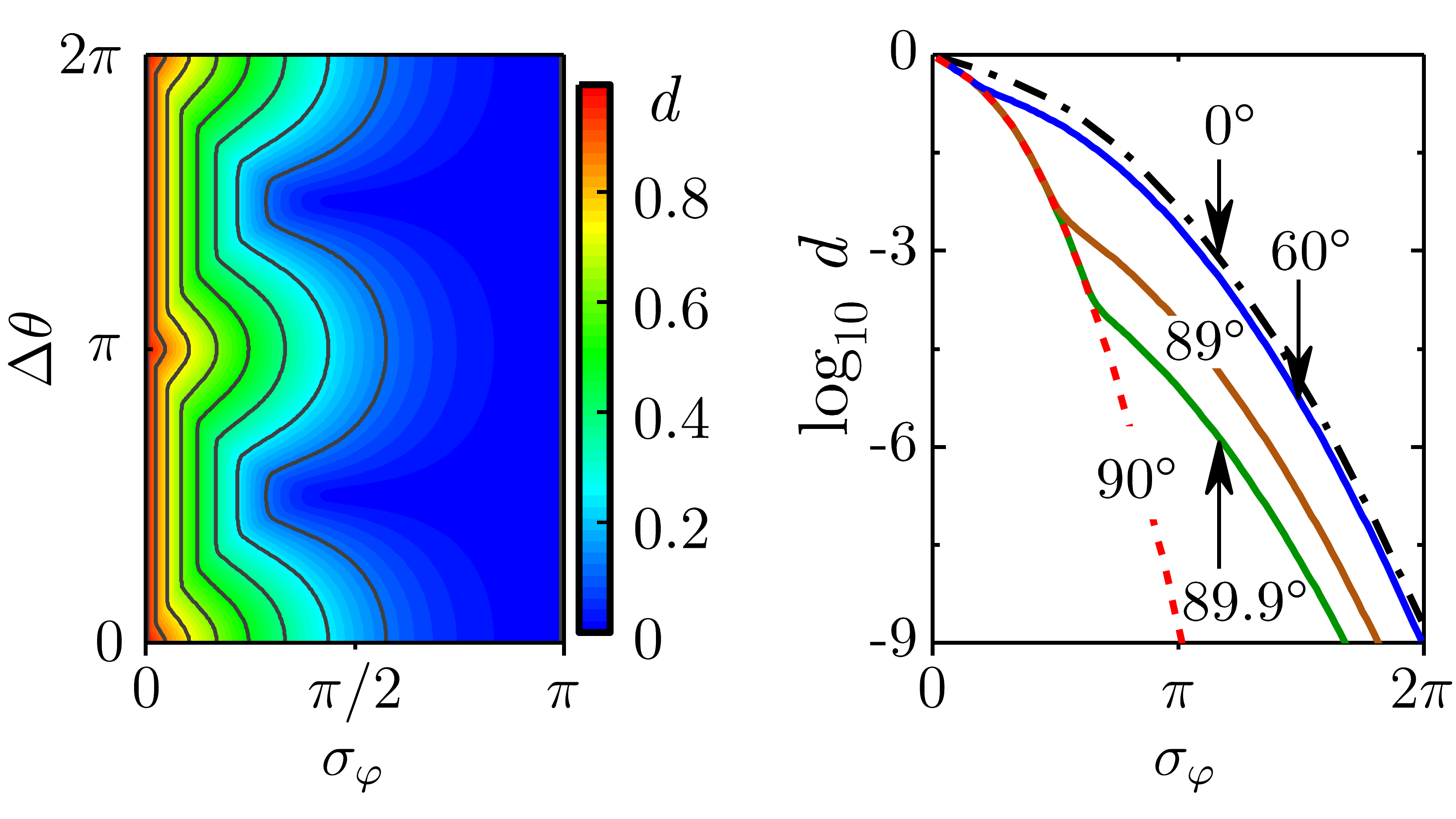}
	\caption{\label{fig:2D} The dependence of the statistical distance $d$ defined by Eq.~\eqref{eq:statisticalDistanceForPDFs2} on $\sigma_\varphi$  and $\Delta\theta$ in the form of a color map (on the left). The slices of the map at different values of $\Delta\theta$  are shown on the right.}
\end{figure}

The dependence of the statistical distance defined by Eq.~\eqref{eq:statisticalDistanceForPDFs2} on $\sigma_\varphi$  and $\Delta\theta$  in the form of a color map is shown in Fig.~\ref{fig:2D} on the left. The slices of the map at different values of $\Delta\theta$  are shown in Fig.~\ref{fig:2D} on the right. First of all, it is clear that the statistical distance is highly sensitive to the value of  $\Delta\theta$. Thus, $d$  is at least 7 orders of magnitude smaller at $\Delta \theta  = \pi (m + {1/2})$  than at  $\Delta \theta  = \pi m$. In fact, one can see from Fig.~\ref{fig:2D} that  $d \approx {10^{ - 9}}$ at $\Delta \theta  = {\pi/2}$  and $d \approx {10^{ - 2}}$  at $\Delta \theta  = 0$  when ${\sigma _\varphi } = \pi$. Another interesting feature is that the $d({\sigma _\varphi })$  curves are not significantly different from each other in the range of $\Delta \theta$  close to  $\pi m$, whereas significant difference between them is clearly seen in the vicinity of $\Delta \theta  = \pi (m + {1/2})$  (compare, e.g., curves at $\Delta \theta  = 90^\circ$  and $\Delta \theta  = 89.9^\circ$  in Fig.~\ref{fig:2D} on the right). It is also seen from Fig.~\ref{fig:2D} that the greatest value of $d$  does not exceed $10^{-9}$  when  ${\sigma _\varphi } > 2\pi$, which confirms that ${f_{\Delta \varphi }}$  can be considered uniform at these values of  $\sigma_\varphi$. On the other hand, $d$  has quite high values in the range of $\sigma_\varphi$  between $\pi$  and  $2\pi$, particularly when  $\Delta\theta$ is close to  $\pi m$. Random numbers obtained at such $d$  cannot be considered enough random in terms of cryptographic security. Therefore, one should additionally improve the randomness of the digitized interference signal when $\pi  < {\sigma _\varphi } < 2\pi$. As far as we know, such an analysis has not been yet carried out in the literature. In the next section, we show how this can be performed in terms of the QRF.

\subsection{Improving randomness in the range $\pi  < {\sigma _\varphi } < 2\pi$} \label{sec:improvingRandomness}
The choice of the statistical distance as a measure of the deviation of $\tilde f_{\tilde S}^Q$  from $f_{\tilde S}^Q$  is also convenient because $d$  is used in the definition of a randomness extractor and also in the formulation of the leftover hash lemma, which can be used to relate $d$  with the QRF. To demonstrate this relation more consistently, let us first provide some necessary definitions.

Recall that the two distributions, $P_X$  and  $P_Y$, are called   $\varepsilon$-close, if their statistical distance defined by Eq.~\eqref{eq:statisticalDistance} does not exceed  $\varepsilon$:  $d\le\varepsilon$. Roughly speaking, one can distinguish   $\varepsilon$-close distributions only with the probability not exceeding  $\varepsilon$, such that $\varepsilon$ may be also considered as an error parameter. If a distribution $P_X$  is   $\varepsilon$-close to the uniform distribution, then $P_X$  is referred to as quasi-uniform. Recall further that a random variable $X$ is called a   $k$-source ($k$  is a some real number), if the min-entropy $H_\infty$ of $X$  does not exceed $k$ or, equivalently, ${P_X}(x) \le {2^{ - k}}$  for any value $x$  of a random variable  $X$. If the discrete random variable $X$  is defined on a set of elementary events representing binary vectors of a length  $n$, then such a set is denoted as  $\Omega  = {\{ 0,1\} ^n}$. Such a random variable is also called the   $(n,k)$-source, if $X$  obeys the inequality  ${H_\infty }(X) \ge {k/n}$, where $H_\infty$  is a min-entropy per bit. Thus, the random signal digitized with an 8-bit analog-to-digital converter (ADC) can be considered as an $(8,k)$-source, if the min-entropy of the obtained binary sequence does not exceed  $k$. Here, the value of $n$ in the definition of the $(n,k)$-source was put to a bit resolution $b$  of an ADC; however, the digitized random signal may be also grouped into much longer binary sequences with arbitrary  $n$, which will be then considered as a random variable on ${\{ 0,1\} ^n}$  with  $n>b$.

Using the above definitions, we may now give a strict definition of a (seeded) randomness extractor. Let us first recall that a \textit{seed} is an additional random binary sequence with (quasi-)uniform distribution. Further, let there be an   $(n,k)$-source $X$ with distribution $P_X$ and a seed having a uniform distribution on  ${\{ 0,1\} ^l}$. Then the function $E({P_X}):{\left\{ {0,1} \right\}^n} \times {\left\{ {0,1} \right\}^l} \to {\left\{ {0,1} \right\}^m}$  is called a $(k,\varepsilon )$  extractor, if the resulting distribution on ${\left\{ {0,1} \right\}^m}$  is   $\varepsilon$-close to uniform. Obviously, the length $l$  of a seed should be quite small, at least it should be shorter than the output, $l<m$, because otherwise the extractor will become trivial, i.e. it would be possible to output the seed itself. 

Seeded randomness extractors are generally implemented via the so-called 2-universal hash-functions, whose effectiveness is guaranteed by the leftover hash lemma (LHL) \cite{Nisan99}. According to this lemma, the transformation ${\left\{ {0,1} \right\}^n} \to {\left\{ {0,1} \right\}^m}$ defined by a set of 2-universal hash-functions and performed on the $(n,k)$-source is a  $(k,\varepsilon)$-extractor, if
\begin{equation}\label{eq:LHL}
m = k - 2{\log _2}({1/\varepsilon }),
\end{equation}
where $\varepsilon$  is the required error parameter. The sense of this theorem can be roughly reformulated as follows. If there is a binary string of length $n$  obtained from a weak entropy source, then one can extract $m$  truly random bits with the use of a set of 2-universal hash-functions, and the upper bound for $m$  is the value of the min-entropy of the raw sequence. Generally, $nH_\infty$  bits could be extracted from the raw sequence with a hypothetical ideal randomness extractor; however, according to the LHL, 2-universal hash-functions allow extracting such number of bits ($m = k = n{H_\infty }$) only with  $\varepsilon=1$, which does not guarantees the uniformity of the output sequence. In other words, the number of truly random bits that can be extracted with the use of 2-universal hash functions is less than  $nH_\infty$; however, we can guarantee the uniformity of the resulting distribution (up to the error parameter $\varepsilon$). In cryptographic applications, $\varepsilon$  is chosen over a wide range of values:  $\varepsilon \sim{10^{ - 10}} - {10^{ - 30}}$.

Instead of the error parameter  $\varepsilon$ one can use the ratio between the lengths of the input and output binary sequences, which is sometimes referred to as a reduction factor:  $\gamma=n/m$. Using the leftover hash lemma, one can easily find the relationship between $\varepsilon$  and  $\gamma$. Indeed, inserting $m=n/\gamma$  into Eq.~\eqref{eq:LHL} we will find:
\begin{equation}
\varepsilon  = {2^{ - nr}},
\end{equation}
where
\begin{equation}
r = \frac{{{{k\gamma }/n} - 1}}{{2\gamma }}.
\end{equation}
In a similar way, we can find the relationship between the QRF and  $\varepsilon$. However, it should be noted here that the raw binary sequence, obtained by digitizing (with a comparator) a random interference signal, is already uniform from the point of view of classical randomness (of course, with the proper choice of the threshold voltage on the comparator). Obviously, such a raw sequence does not require additional post-processing; therefore, the LHL is not applicable here in the usual sense. That is why the definition of the QRF dispenses with LHL. Nevertheless, we can formally insert $m=n/\Gamma$  into Eq.~\eqref{eq:LHL} taking also into account that we are interested in quantum entropy, so that we should write  $k = nH_\infty ^Q = n$, which yields
\begin{equation}\label{eq:LHLwithGamma}
\frac{n}{\Gamma } = n - 2{\log _2}({1/{{\varepsilon _c}}}),
\end{equation}
where we have introduced an effective error parameter  $\varepsilon_c$, which is related with the contribution of classical noise and not with the non-uniformity of the raw binary sequence. One can easily find from Eq.~\eqref{eq:LHLwithGamma} that
\begin{equation}\label{eq:epsilonC}
{\varepsilon _c} = {2^{ - nR}},
\end{equation}
where
\begin{equation}
R = \frac{{\Gamma  - 1}}{{2\Gamma }}.
\end{equation}

Continuing to develop this approach, we can use Eq.~\eqref{eq:LHLwithGamma} to solve the problem indicated in the title of this section. First, we note that to consider the classical contribution from $\Delta\theta$  in  $\tilde{S}$, it is necessary to increase the value of the QRF introducing a modified factor  $\tilde{\Gamma}$, such that $\tilde \Gamma  > \Gamma$  at ${\sigma _\varphi } \in [\pi ,2\pi ]$  and $\tilde \Gamma  = \Gamma $  at  ${\sigma _\varphi } > 2\pi$. As was shown in the previous section, the statistical distance between ${f_{\Delta \varphi }}$  and the uniform distribution exhibits maximum when  $\Delta \theta  = \pi m$ (let us denote the corresponding value of $d$  as  ${d_{\pi m}}({\sigma _\varphi })$); therefore, it makes sense to use ${d_{\pi m}}({\sigma _\varphi })$  as a parameter characterizing the non-uniformity of  ${f_{\Delta \varphi }}$. Since we neglect the non-uniformity of ${f_{\Delta \varphi }}$  at  ${\sigma _\varphi } > 2\pi $, it seems appropriate to determine the error parameter as
\begin{equation}\label{eq:epsilonQ}
{\varepsilon _Q} \equiv {\varepsilon _Q}({\sigma _\varphi }) = \frac{{{d_{\pi m}}(2\pi )}}{{{d_{\pi m}}({\sigma _\varphi })}}.
\end{equation}
The modified quantum reduction factor $\tilde{\Gamma}$  can be then defined as follows:
\begin{equation}
\frac{n}{{\tilde \Gamma }} = n - 2{\log _2}({1/{{\varepsilon _c}}}) - 2{\log _2}({1/{{\varepsilon _Q}}}),
\end{equation}
whence, using Eq.~\eqref{eq:epsilonC}, we may find the relation between  $\tilde{\Gamma}$ and  $\Gamma$:
\begin{equation}\label{eq:gammaTilde}
\tilde \Gamma ({\sigma _\varphi }) = \frac{{n\Gamma }}{{n - 2\Gamma {{\log }_2}[{1/{{\varepsilon _Q}}}]}}.
\end{equation}
One can see from Eqs.~\eqref{eq:epsilonQ} and \eqref{eq:gammaTilde} that $\varepsilon_Q=1$  at  $\sigma_\varphi=2\pi$, so $\tilde{\Gamma}$  becomes equal to  $\Gamma$. We may thus write for QRF: 
\begin{equation}\label{eq:QRF}
\text{QRF} = \left\{ {\begin{array}{*{20}{l}}
	{\infty ,\,\,\,{\sigma _\varphi } < \pi;}\\
	{\tilde \Gamma ,\,\,\,{\sigma _\varphi } \in [\pi ,2\pi ];}\\
	{\Gamma ,\,\,\,{\sigma _\varphi } > 2\pi .}
	\end{array}} \right.
\end{equation}

\subsection{Stochastic rate equations}
A fundamental noise source in the output of a laser is the quantum shot noise due to the random electron transitions producing spontaneous emission events \cite{Henry82,Haug69,McCumber66,Morgan72,Henry86}. Probability properties of spontaneous transitions are determined by their relation to zero-point (vacuum) fluctuations of the electromagnetic field \cite{Loudon,Glauber2006}, which are generally considered to be perfectly uncorrelated and broadband. Therefore phase fluctuations in a semiconductor laser are generally assumed to have the same properties. It is sometimes argued that the description of quantum noise in a semiconductor laser should be performed with the method developed by M. Lax \cite{Lax66,Lax67}. Namely, one should use a Markovian model for a set of atoms and a radiation field interacting with some reservoirs, which are defined mathematically by a set of damping constants and a set of non-commuting noise operators (quantum Langevin forces). In many cases, however, it is preferable to employ the simple model developed by C. Henry \cite{Henry82}, who showed how the spontaneous emission should be incorporated into the classical field rate equation for a semiconductor laser, such that the drift and diffusion coefficients for the field intensity and the phase remained consistent with the fully quantum description. For our purposes, a standard system of laser rate equations with classical Langevin terms \cite{Petermann,Henry82,Henry86,Tartwijk95,AgrawalDutta} is well-suited; therefore, we will use here this approach. (See Appendix \ref{sec:derivationOfStochactisEquations} for more details on stochastic rate equations we use here for simulations.)

\subsection{Phase diffusion: Monte-Carlo simulations}\label{sec:monteCarloSimulations}
\begin{figure}[t]
	\includegraphics[width=\columnwidth]{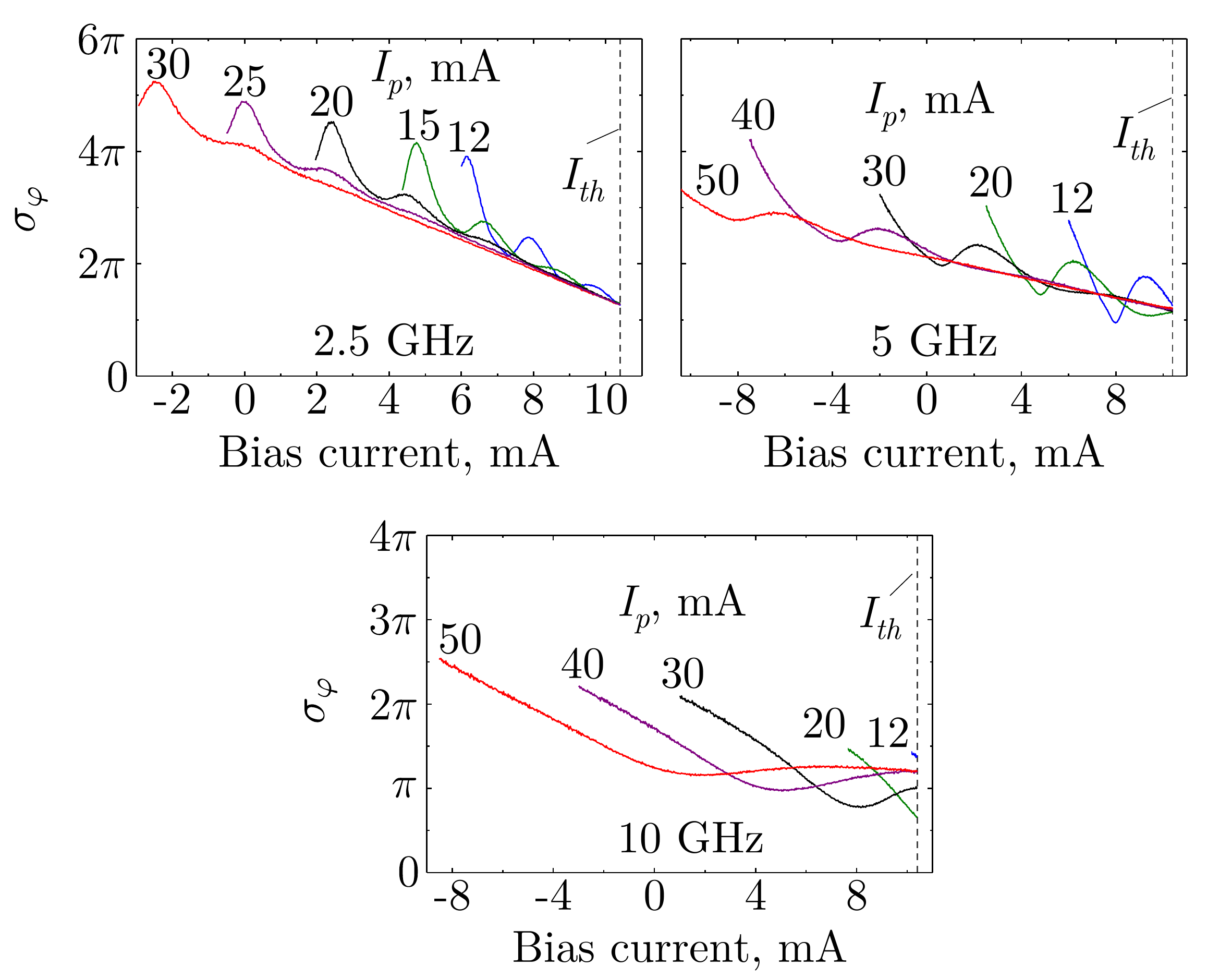}
	\caption{\label{fig:CUDAsimulations} Theoretical dependences of the phase diffusion standard deviation ($\sigma_{\varphi}$) on the bias current $I_b$ and modulation current amplitude $I_p$ at different pulse repetition rates.}
\end{figure}

For the numerical study of the phase diffusion between the pulses of a single-mode gain-switched semiconductor laser, we performed Monte-Carlo simulations of the phase evolution between adjacent optical pulses using Eq.~\eqref{eq:differenceSheme}. The pump current was assumed to have a form of a square wave, $I(t) = {I_b} + {I_p}(t)$, where $I_b$ is the bias current and  $I_p(t)$ changed abruptly from 0 to  $I_p$. We performed integration of Eq.~\eqref{eq:differenceSheme} from 0 to time $T_p$  corresponding to a period of pulse repetition, i.e. the phase was allowed to diffuse during the time between two adjacent pulses. Initial conditions ($N_0$, $Q_0$  and  $\varphi_0$) were chosen so that there were no transients. After each such integration, we obtained a value  $\varphi ({T_p})$ of a random phase, which exhibited Gaussian distribution, whose standard deviation we have denoted as  ${\sigma _\varphi }({T_p}) \equiv {\sigma _\varphi }$. 50 000 iterations (random values of  $\varphi ({T_p})$) were found to be enough to find a value of $\sigma_\varphi$  for given values of  $I_b$, $I_p$   and a given set of laser parameters. In Fig.~\ref{fig:CUDAsimulations}, we presented the dependence of $\sigma_\varphi$  on the bias current $I_b$  at different values of the pulse repetition rate  $f_p=1/T_p$; we used three different values of  $f_p$: 2.5~GHz, 5~GHz, 10~GHz. At each pulse repetition rate, we calculated several curves corresponding to different values of  $I_p$. For each  $I_p$, we chose the range of the bias current variation from the value $I_b^{min}$  corresponding to stable pulsation at a given $I_p$  up to the threshold value defined by  $I_{th}=N_{th}e/\tau_e$. The gain compression factor  $\gamma_P$ in Fig.~\ref{fig:CUDAsimulations} was put to 20~W$^{-1}$. Other laser parameters used in simulations are listed in Table~\ref{tab:laserParameters}. In order to achieve high performance and reduce the computation time, we performed computations on a Compute Unified Device Architecture (CUDA) platform with the Nvidia video card equipped by a CUDA-powered graphics processing unit.
\begin{table}[b]
	\caption{\label{tab:laserParameters}%
		Laser parameters used in simulations.
	}
	\begin{ruledtabular}
		\begin{tabular}{lc}
			\textrm{Parameter}&
			\textrm{Value}\\
			\colrule
			Photon lifetime  $\tau_{ph}$, ps & 1.0 \\
			Electron lifetime  $\tau_e$, ns & 1.0\\
			Quantum differential output $\eta$ & 0.3 \\
			Transparency carrier number  $N_{tr}$ &$6.0\times10^7$\\
			Threshold carrier number  $N_{th}$ & $6.5\times10^7$\\
			Spontaneous emission coupling factor  $C_{sp}$ & $10^{-5}$\\
			Confinement factor $\Gamma$ & 0.12\\
			Linewidth enhancement factor $\alpha$ & 6\\
			Central lasing frequency  $\omega_0/(2\pi)$, THz & 193.548			
		\end{tabular}
	\end{ruledtabular}
\end{table}

One can see from Fig.~\ref{fig:CUDAsimulations} that $\sigma_\varphi$  increases towards lower values of the bias current. This reflects the fact that the number of carriers decreases faster at a lower pump current between laser pulses, which leads to a faster decrease of the field intensity inside the laser cavity and, consequently, to a faster decoherence due to spontaneous emission. An interesting feature here is a non-monotonic behavior of $\sigma_\varphi(I_b)$  curves, which exhibit "damped oscillations". In section \ref{sec:discussion}, we will discuss this result and provide an explanation of these "oscillations".

\section{Experiment}\label{sec:experiment}
\subsection{Experimental setup description}\label{sec:experimentalSetup}
An obvious way to measure the phase diffusion between pulses of a gain-switched laser is to measure the pulse interference using an unbalanced Mach-Zehnder interferometer (uMZI), whose time delay is chosen to be equal to the laser pulse repetition period. It should be noted that it is difficult to use a fiber-optic interferometer for phase diffusion measurements due to the temperature drift in the fiber. In fact, due to its relatively large size, it is quite problematic to stabilize it in temperature; therefore, thermal fluctuations may introduce significant errors to the measured values of  $\sigma_\varphi$, especially if a measurement takes a long time. Also, it is difficult to make the fiber-optic delay line that would precisely correspond to a predetermined value of the pulse repetition rate using just a fiber fusion splicer. Because of this, we used an integrated uMZI, where it is quite simple to implement active temperature stabilization and to perform fine adjustment of the phase difference between the interferometer arms.

\begin{figure}[b]
	\includegraphics[width=\columnwidth]{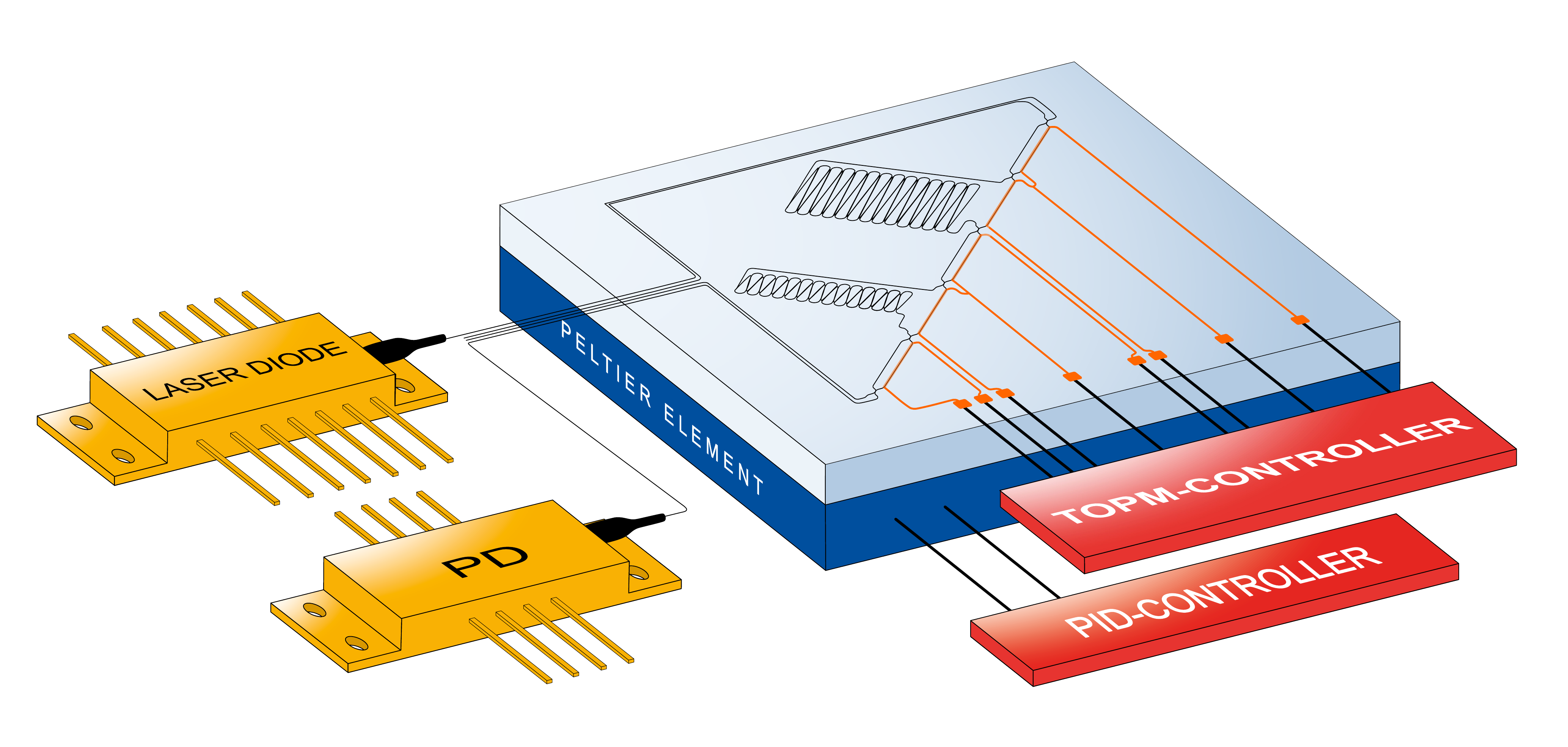}
	\caption{\label{fig:uMZI} Schematic representation of the experimental setup to measure phase diffusion in a gain-switched laser.}
\end{figure}

Schematic representation of the experimental setup used to measure phase diffusion between laser pulses is shown in Fig.~\ref{fig:uMZI}. We employed the 1550~nm distributed feedback (DFB) laser module (Gooch~\&~Housego, AA0701) of 12 Gbps modulation bandwidth. The bias current was controlled by the high-stability laboratory power supply, whereas modulation signals of 2.5 and 1.25 GHz were generated by a phase-locked loops followed by a broadband amplifier. The optical signal was detected with the InGaAs fixed gain amplified detector (Thorlabs, PDA8GS) of 9.5 GHz bandwidth (PD in Fig.~\ref{fig:uMZI}), and the signal processing was performed using the Teledyne Lecroy digital oscilloscope (WaveMaster 808Zi-A) with 8 GHz bandwidth and temporal resolution of 25 ps.

The laser diode with the polarization-maintaining fiber output was coupled to the photonic integrated circuit (PIC) containing a cascade of uMZIs. For simplicity,  we presented in Fig.~\ref{fig:uMZI} only 2 uMZIs with delay lines of 400 and 800 ps. The PIC contained also additional balanced MZIs (3 of them are shown in Fig.~\ref{fig:uMZI}), which were used to control the splitting ratios of the interferometers, as well as to choose the uMZI with the desired delay line. Each MZI on the chip was equipped by a thermo-optical phase modulator (TOPM) in the form of a resistive heater (a metal band) deposited over a waveguide corresponding to one of the interferometer arms. Heaters were controlled via a multi-channel digital-to-analog converter (DAC) indicated as a TOPM-controller in Fig.~\ref{fig:uMZI}. To set the desired configuration of DAC voltages, we applied to the laser a low-frequency (312.5 MHz) modulation signal, which yielded in a train of short laser pulses with the repetition period of 3.2 ns. We then set up the controller so that each of these pulses was split at the output of the chip into a pair of pulses of the same intensity. The time delay between the pulses in a pair was set either to 400 ps or to 800 ps depending on the choice of the delay line. (Some details concerning the control over delay lines are given in Appendix~\ref{app:interferometerControl}.)

\begin{figure}[t]
	\includegraphics[width=\columnwidth]{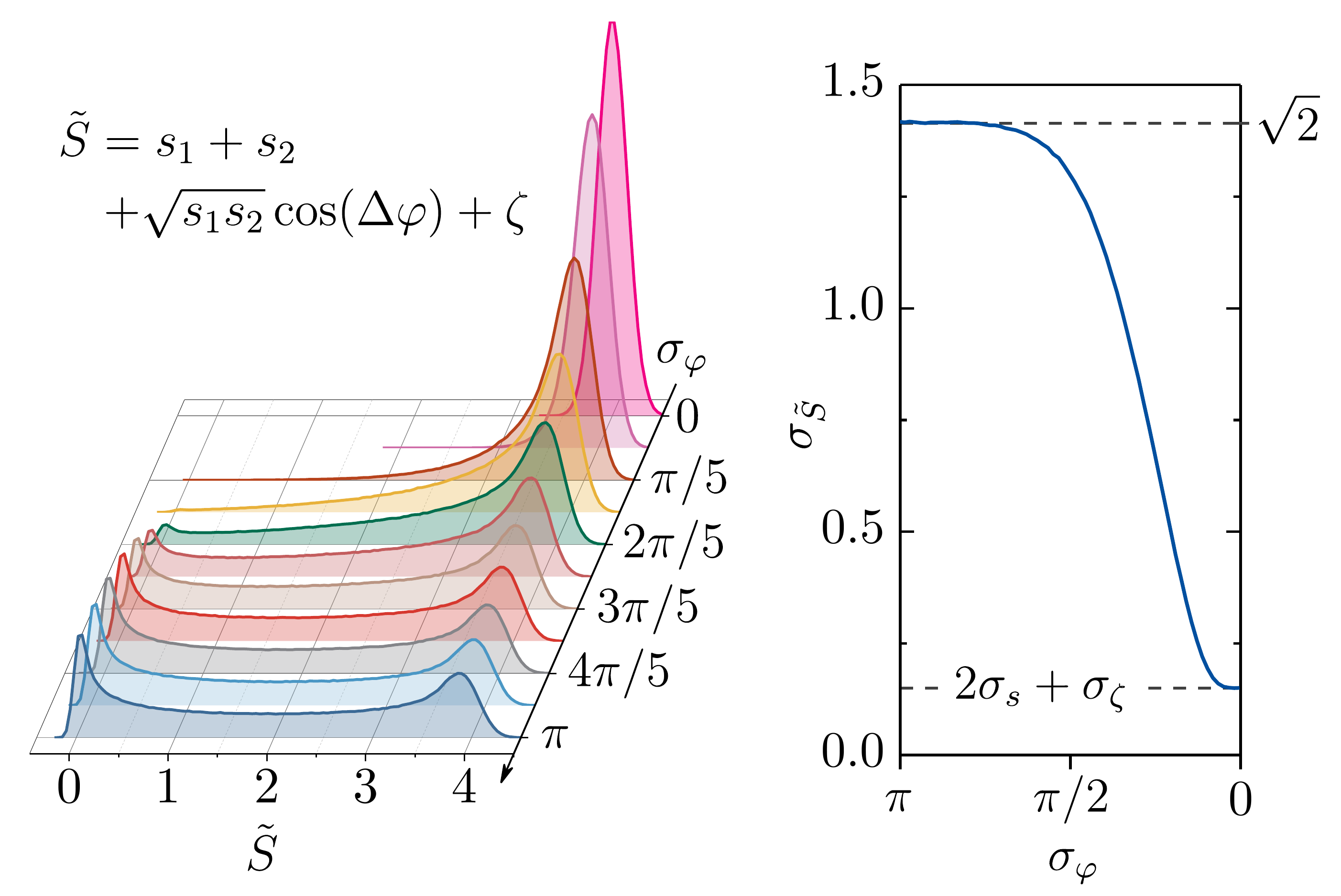}
	\caption{\label{fig:PDFsTheoretical} Theoretical dependence of the probability density function of the normalized interference signal $\tilde{S}$ (Eq.~\eqref{eq:resultOfInterference8}) on the standard deviation of phase fluctuations $\sigma_\varphi$ (on the left) and the corresponding dependence of its standard deviation $\sigma_{\tilde{S}}$ (on the right).}
\end{figure}

To maintain a stable temperature, the chip was mounted on the Peltier thermoelectric cooler controlled by a commercially available temperature controller (PID-controller in Fig.~\ref{fig:uMZI}). Note that the Peltier element can be used along with TOPMs to control the phase difference between the interferometer arms; moreover, the change in the chip temperature allowed for much greater modulation depth than resistive heaters, although the latter allowed for much faster phase modulation (up to several kilohertz). Since for our experiments there was no need to change the phase with such a frequency, we used the Peltier element to vary the phase in the interferometer arms.

\subsection{Phase diffusion measurements}
The main object of measurements in our experiments was the probability density function of laser pulse interference $f_{\tilde{S}}$. Before proceeding to the study of experimental PDFs, it seems appropriate to simulate the interference statistics. For simulations, we used Eq.~\eqref{eq:resultOfInterference8} with $\eta=1$, and random amplitudes $s_1$, $s_2$ with Gaussian PDF: $\bar{s}_1=\bar{s}_2=1$ and $\sigma_s=0.05$. The same value of the standard deviation was assumed for the additive Gaussian noise: $\sigma_\zeta=0.05$. We varied the standard deviation of phase diffusion $\sigma_\varphi$ from $\pi$ to 0 with a step of $\pi/10$. The theoretical evolution of the interference PDF with such a change of $\sigma_\varphi$ is shown in Fig.~\ref{fig:PDFsTheoretical} on the left. One can see that when $\sigma_\varphi$ is close to $\pi$, the PDF has two pronounced maxima near $\tilde{S}=0$ and $\tilde{S}=4$ with the left maximum noticeably thinner and higher than the right one. When decreasing $\sigma_\varphi$ (and assuming that the phase change $\Delta\theta$ in the interferometer is zero), the right maximum of the PDF starts to grow, and when $\sigma_\varphi$ tends to zero, the PDF turns into Gaussian curve with standard deviation equal to $2{\sigma _s} + {\sigma _\zeta }$. Corresponding evolution of $\sigma_{\tilde{S}}$ (standard deviation of $f_{\tilde{S}}$) is shown in Fig.~\ref{fig:PDFsTheoretical} on the right: when $\sigma_\varphi>\pi$, $\sigma_{\tilde{S}}$ tends to the value $\sigma_{\tilde{S}}=\sqrt{2}$, which corresponds to the standard deviation of the quantum PDF $f_{\tilde{S}}^Q$ defined by Eq.~\eqref{eq:quantumPDF}.

\subsubsection{Experimental PDFs}

\begin{figure}[t]
	\includegraphics[width=\columnwidth]{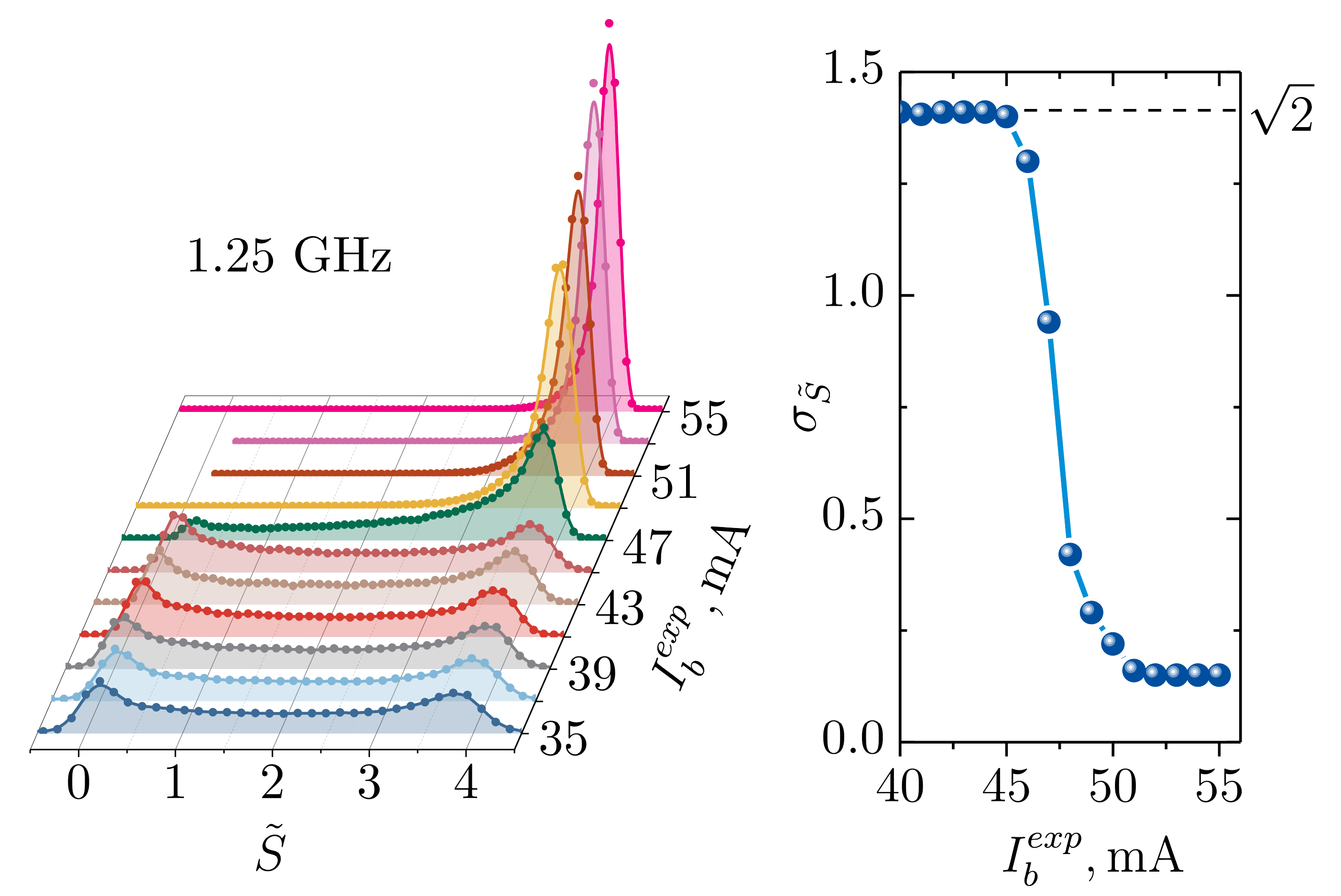}
	\caption{\label{fig:PDFsExperimental1250} Experimental dependence of the probability density function of the normalized interference signal $\tilde{S}$ on the bias current $I_b^{exp}$ (on the left) and the corresponding dependence of its standard deviation $\sigma_{\tilde{S}}$ (on the right) at pulse repetition frequency $f_p=1.25$~GHz.}
\end{figure}

\begin{figure}[b]
	\includegraphics[width=\columnwidth]{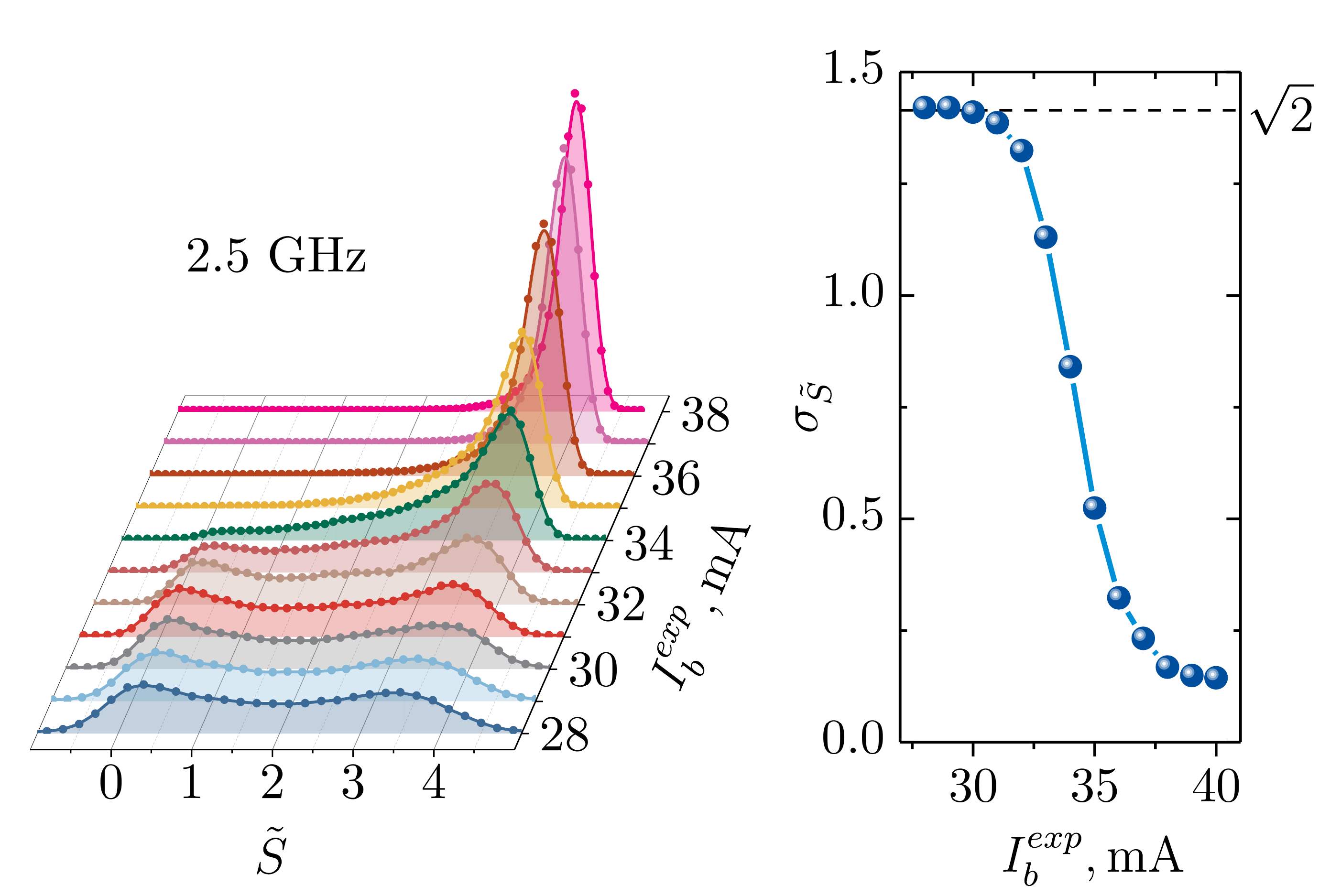}
	\caption{\label{fig:PDFsExperimental2500} Experimental dependence of the probability density function of the normalized interference signal $\tilde{S}$ on the bias current $I_b^{exp}$ (on the left) and the corresponding dependence of its standard deviation $\sigma_{\tilde{S}}$ (on the right) at pulse repetition frequency $f_p=2.5$~GHz.}
\end{figure}

Figures~\ref{fig:PDFsExperimental1250} and \ref{fig:PDFsExperimental2500} (on the left) show experimental statistics of the normalized interference signal, which were recorded as histograms with the oscilloscope at a laser pulse repetition rate of 1.25 and 2.5~GHz, respectively, at various bias currents $I_b^{exp}$ (the procedure of normalization is explained in Appendix \ref{app:signalNormalization}). As was shown in section \ref{sec:monteCarloSimulations} (see Fig.~\ref{fig:CUDAsimulations}), at sufficiently high values of the modulation current ($I_p>30$~mA) and not very high values of the pulse repetition frequency ($f_p<5$~GHz), $\sigma_\varphi$ below the $2\pi$-value decreases linearly when increasing the bias current $I_b$. So, we may consider that the linear increase of the bias current in the experiment is equivalent to a linear decrease in $\sigma_\varphi$. 

One can see that experimental PDFs repeat the evolution shown in Fig.~\ref{fig:PDFsTheoretical}. Note, however, that experimental statistics (particularly in case of short optical pulses without spectral filtering) is generally affected by the "chirp~+~jitter" effect \cite{Shakhovoy2021}. Therefore, to achieve a good correspondence between theoretical and experimental PDFs, we put the optical filter between the interferometer output and the photodetector. (We used the Santec OTF-980 optical tunable filter; the bandpass was put to 6.25 GHz.) 

\begin{figure}[t]
	\includegraphics[width=\columnwidth]{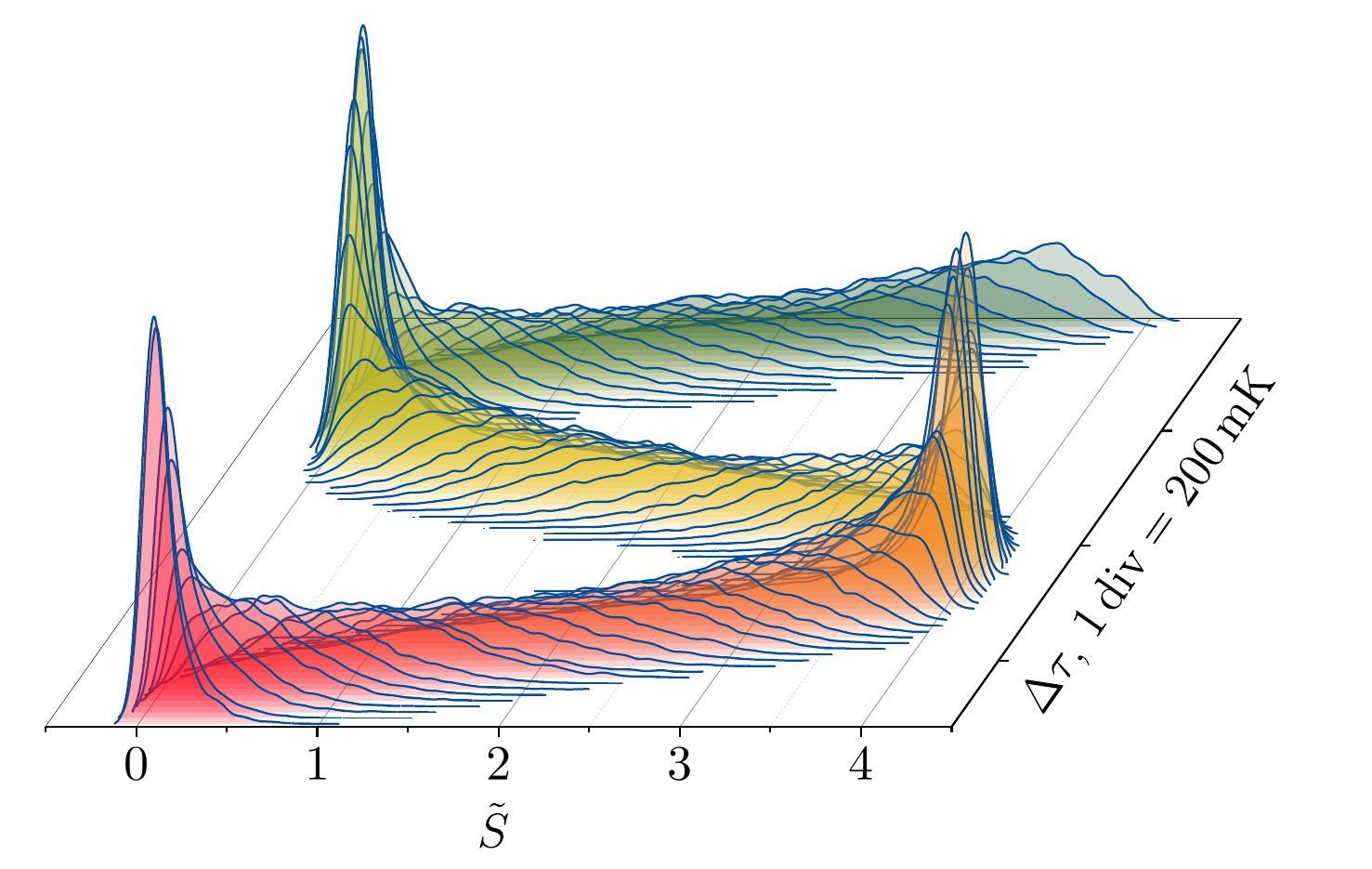}
	\caption{\label{fig:fringesVsTemperature} An example of measured probability density function of the laser pulse interference as a function of the interferometer temperature. Pulse repetition rate was 1.25~GHz; the bias current was $I_b^{exp}=54$~mA. The temperature shift $\Delta \tau$ was measured with respect to 300~K.}
\end{figure}

\begin{figure}[t]
	\includegraphics[width=\columnwidth]{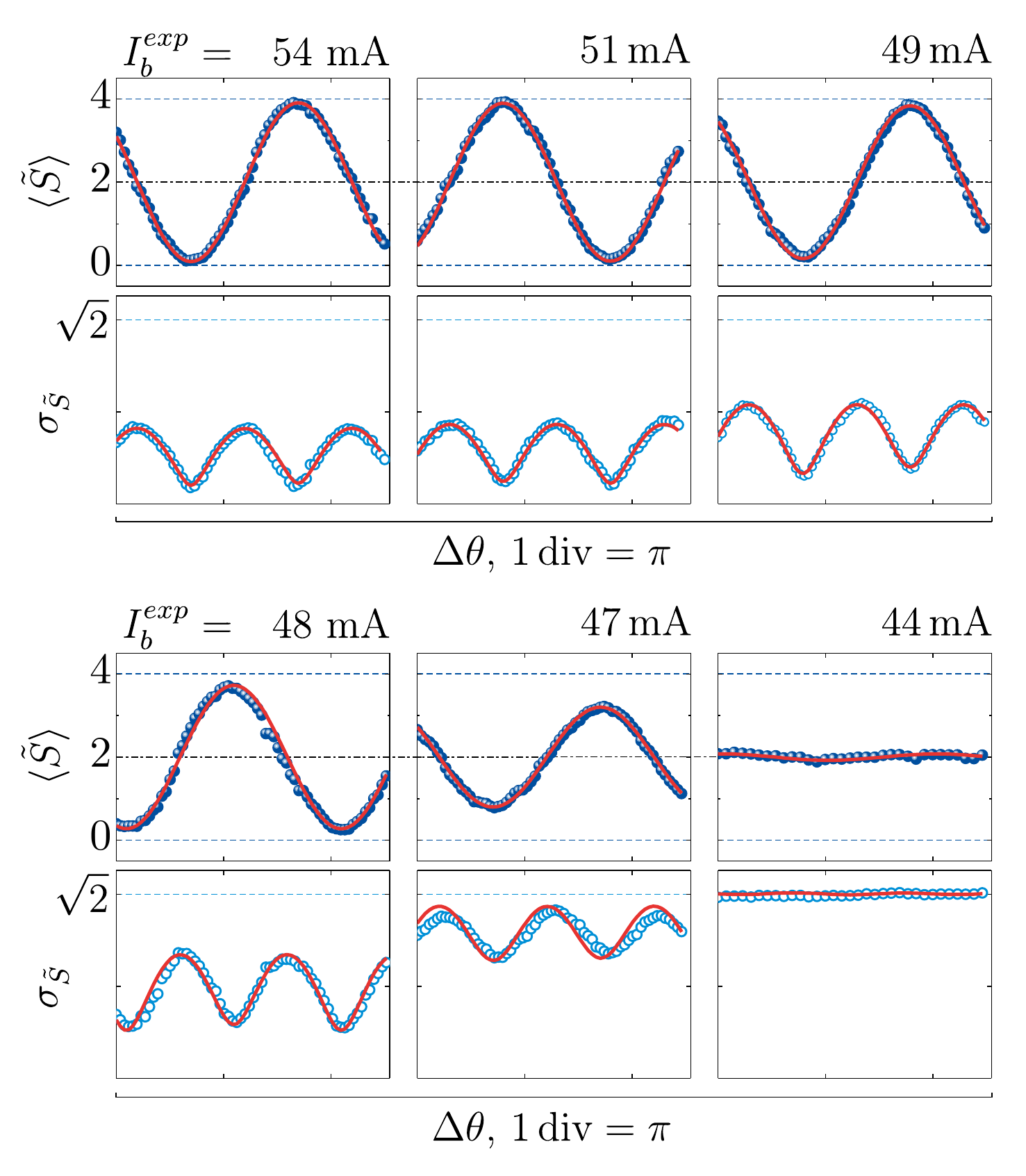}
	\caption{\label{fig:fringes1250} Experimental dependences of the mean value $\langle \tilde{S}\rangle$ of the interference signal on the phase shift $\Delta\theta$ (filled circles) and corresponding statistical interference fringes $\sigma_{\tilde{S}}(\Delta\theta)$ (empty circles) at pulse repetition frequency of 1.25~GHz.}
\end{figure}

\begin{figure}[t]
	\includegraphics[width=\columnwidth]{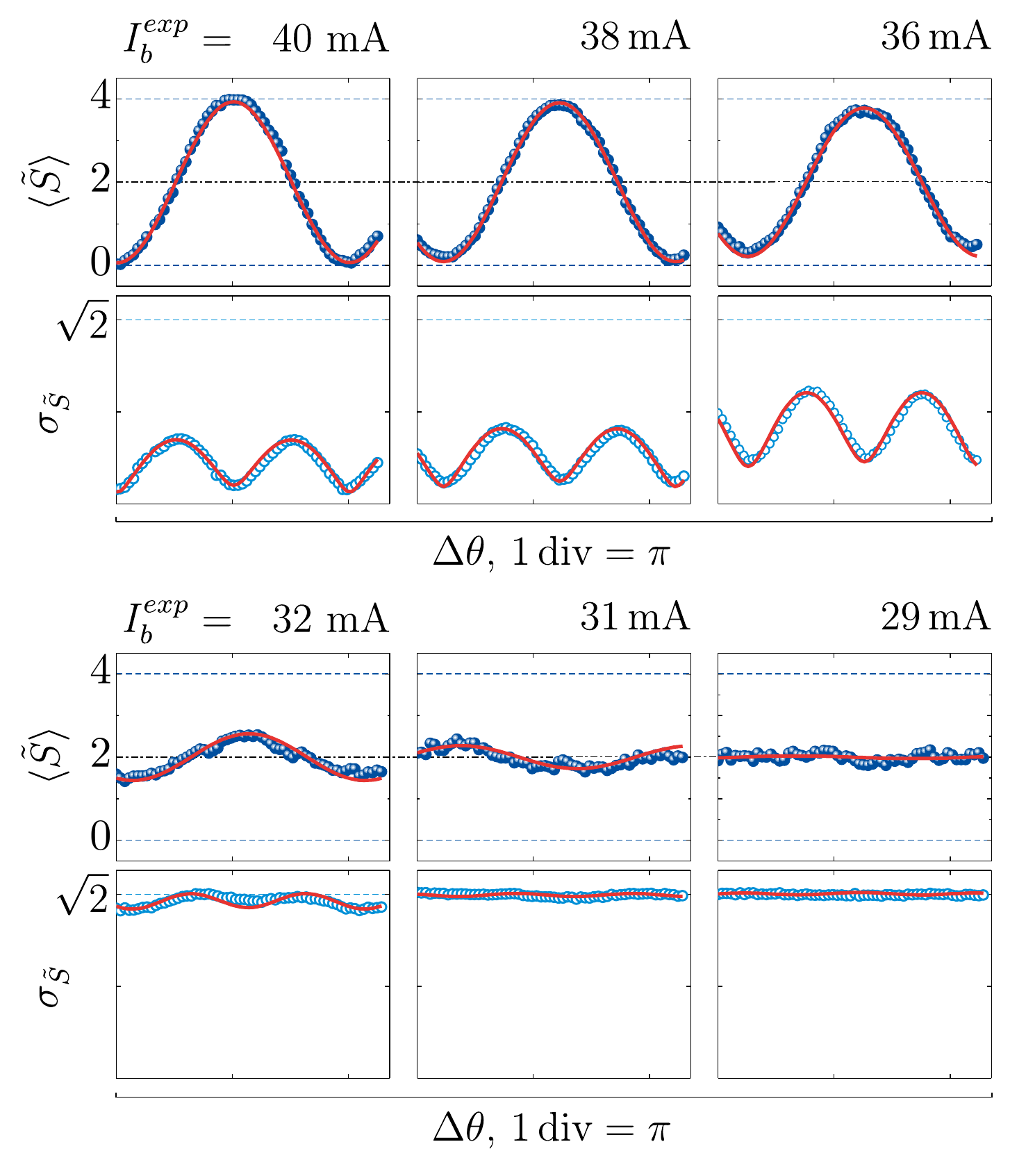}
	\caption{\label{fig:fringes2500} Experimental dependences of the mean value $\langle \tilde{S}\rangle$ of the interference signal on the phase shift $\Delta\theta$ (filled circles) and corresponding statistical interference fringes $\sigma_{\tilde{S}}(\Delta\theta)$ (empty circles) at pulse repetition frequency of 2.5~GHz.}
\end{figure}

When performing simulations in Fig.~\ref{fig:CUDAsimulations}, it was convenient to define the bias current $I_b$ as the minimum value of the pump current. (The peak-to-peak value $I_p$ of the modulation current was "measured" from $I_b$.) However, in the experiment, it was convenient to define the pump current as ${I^{exp}}(t) = I_b^{exp} + I_p^{exp}(t)$, where $I_p^{exp}(t)$ is changed from $-I_p^{exp}/2$ to $I_p^{exp}/2$. In this case, the minimum value of the pump current is $I_b^{exp} - {{I_p^{exp}}/2}$. In Fig.~\ref{fig:PDFsExperimental1250}, the pump current axes are plotted in terms of $I_b^{exp}$.

The dependences of the standard deviation $\sigma_{\tilde{S}}$ of the normalized signal $\tilde{S}$ on the bias current at $f_p=1.25$ and $f_p=2.5$~GHz are shown in Figs.~\ref{fig:PDFsExperimental1250} and \ref{fig:PDFsExperimental2500} on the right. (For each value of $I_b^{exp}$ we set $\Delta\theta=0$ by adjusting the temperature of the interferometer.) One can see that the experimental dependences $\sigma_{\tilde{S}}(I_b^{exp})$ are in a good agreement with the theoretical dependence ${\sigma _{\tilde S}}({\sigma _\varphi })$ shown in Fig.~\ref{fig:PDFsTheoretical}. Comparing theoretical dependence ${\sigma _{\tilde S}}({\sigma _\varphi })$ in Fig.~\ref{fig:PDFsTheoretical} with experimental dependences $\sigma_{\tilde{S}}(I_b^{exp})$ in Figs.~\ref{fig:PDFsExperimental1250} and \ref{fig:PDFsExperimental2500}, one can easily estimate (just by eye) what value of the bias current approximately corresponds to $\sigma_\varphi=\pi/2$ (at this value the curve $\sigma_{\tilde{S}}(I_b^{exp})$ starts to bend) and to $\sigma_\varphi=\pi$ (at this value the curve $\sigma_{\tilde{S}}(I_b^{exp})$ reaches the maximum).

We can also employ the $\sigma_{\tilde{S}}(I_b^{exp})$ curves to estimate the peak-to-peak value of the modulation current. Thus, it is quite reasonable to assume that the bend in the curve occurs when $I_b\approx I_{th}$; therefore, for the corresponding value of $I_b^{exp}$ we may write the simple equation: $I_b^{exp}-I_p^{exp}/2=I_{th}$. The threshold current $I_{th}$ of our laser was approximately 10~mA, so, at 1.25~GHz we had $I_p^{exp}\approx70$~mA, whereas at 2.5~GHz we had $I_p^{exp}\approx40$~mA.

\subsubsection{Interference fringes}
A more accurate estimate of $\sigma_\varphi$ can be obtained by analyzing conventional interference fringes (the dependences of the mean value $\langle \tilde{S}\rangle$ on the phase change $\Delta\theta$ in the interferometer), and also by investigating the curves $\sigma_{\tilde{S}}(\Delta\theta)$, which we refer here to as $\sigma_{\tilde{S}}$-fringes or \textit{statistical} interference fringes. To measure fringes at a given value of the bias current, we varied the temperature of the interferometer (with the step of 10~mK for the delay line of 800~ps and with the step of 20~mK for the delay line of 400~ps) and for each temperature recorded the PDF. The result of such measurements at $f_p=1.25$~GHz for $I_b^{exp}=54$~mA is shown in Fig.~\ref{fig:fringesVsTemperature} as an example. Experimental fringes extracted from these data at some selected values of the bias current are shown in Fig.~\ref{fig:fringes1250} for $f_p=1.25$~GHz and in Fig.~\ref{fig:fringes2500} for $f_p=2.5$~GHz.
 
It can be easily shown \cite{Kobayashi2014,Eichen84} that standard deviation of the phase diffusion $\sigma_\varphi$ can be calculated via the relation $\sigma_\varphi=\sqrt{-2\ln\eta}$, where $\eta$ is the visibility of the pulse interference. Visibility, in turn, can be obtained by fitting interference fringes (filled circles in Figs.~\ref{fig:fringes1250} and \ref{fig:fringes2500}) with the formula $\langle \tilde{S}\rangle=2[1+2\eta\sin(\Delta\theta+\varphi_0)]$, where $\varphi_0$ is an auxiliary fitting parameter, which allows taking into account the "initial phase" of the fringe. However, we will use here another approach, namely, we will perform the joint fit of both $\langle \tilde{S}\rangle$ and $\sigma_{\tilde{S}}$ dependences on $\Delta\theta$ with the use of Eqs.~\eqref{eq:resultOfInterference8} and \eqref{eq:gaussianPDF}. The main advantage of such an approach is that it allows determining not only $\sigma_\varphi$, but also $\sigma_s$ (standard deviation of normalized laser pulse intensity fluctuations). In the context of a QRNG, this is extremely useful because intensity fluctuations are generally treated as classical noise, which should be properly taken into account when extracting quantum noise from the interference of laser pulses. The result of the joint fit for each pair of curves is shown in Figs.~\ref{fig:fringes1250} and \ref{fig:fringes2500} by solid red lines.

\begin{figure}[b]
	\includegraphics[width=\columnwidth]{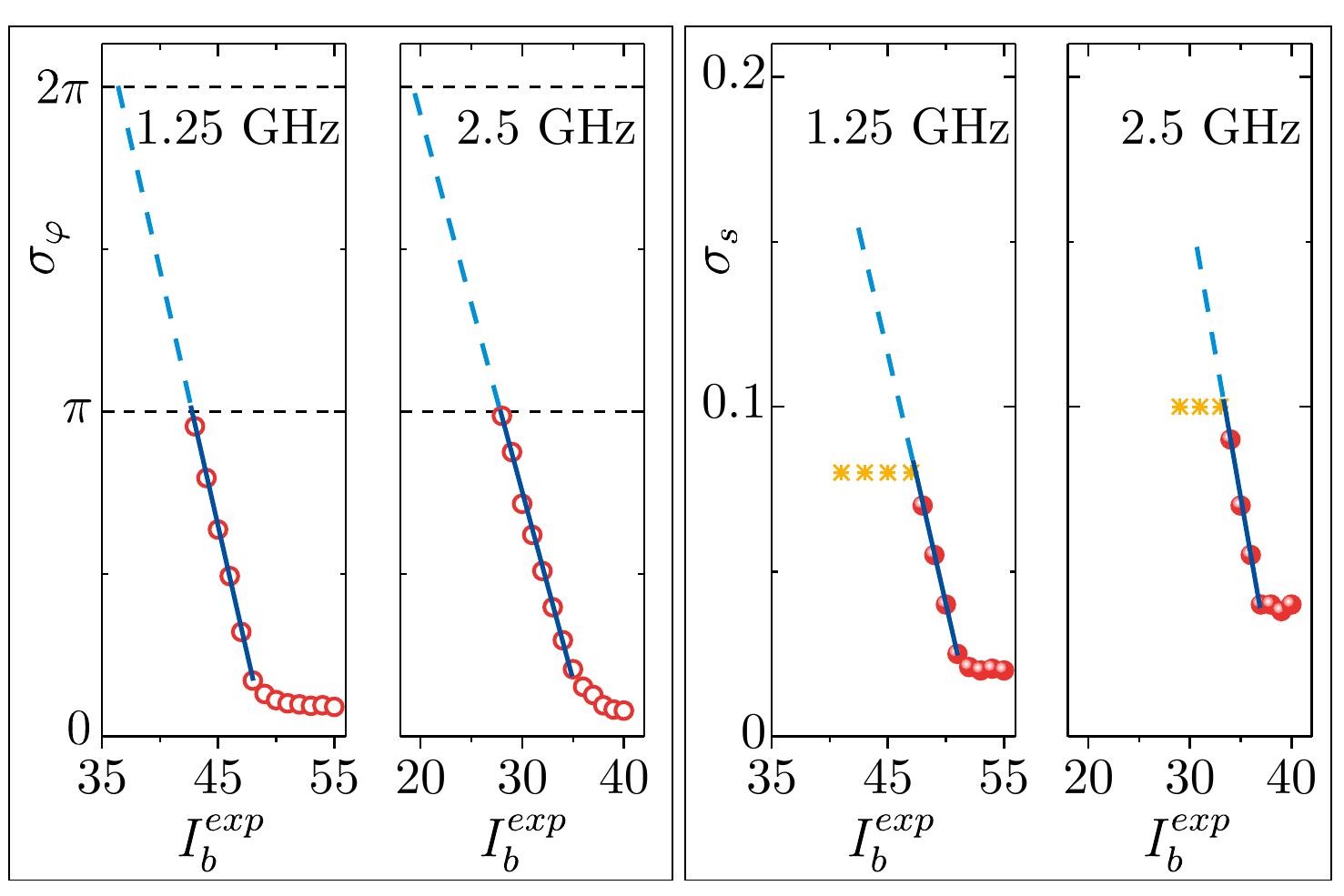}
	\caption{\label{fig:sigmaPhiSigmas} Experimentally measured standard deviations of phase diffusion ($\sigma_\varphi$) and normalized pulse intensity fluctuations ($\sigma_s$) as functions of the bias current at 1.25 and 2.5~GHz of pulse repetition rate. Orange asterisks denote values of $\sigma_s$ that were used when fitting $\sigma_{\tilde{S}}$-fringes at $I_b^{exp}<47$~mA (for 1.25~GHz) and $I_b^{exp}<33$~mA (for 2.5~GHz).}
\end{figure}

Statistical interference fringes differ significantly from the sinusoid and cannot be generally fitted as conventional interference fringes. Particularly, they may exhibit different depth of a dip at constructive and distructive interference, which is clearly seen in some experimental curves (see, e.g., the curve in Fig.~\ref{fig:fringes1250} at $I_b^{exp}=49$~mA). (To demonstrate that this is not related to experimental imperfections, we plotted some theoretical statistical fringes in Fig.~\ref{fig:statisticalFringesTheoretical} of Appendix~\ref{app:statisticalFringes}.)

The results obtained by fitting the curves in Figs.~\ref{fig:fringes1250} and \ref{fig:fringes2500} are presented in Fig.~\ref{fig:sigmaPhiSigmas}. As was discussed above, experimental PDFs at various values of $\sigma_\varphi$ (or rather at various values of $I_b^{exp}$) are almost indistinguishable when $\sigma_\varphi>\pi$; therefore, we may measure the standard deviation of phase diffusion with acceptable precision only when $\sigma_\varphi<\pi$. However, assuming linear evolution of $\sigma_\varphi$ and $\sigma_s$ with $I_b^{exp}$ (at least in the range from $\pi$ to $2\pi$), we may always extrapolate these dependences as shown in Fig.~\ref{fig:sigmaPhiSigmas} by dashed lines.


\section{Discussion}\label{sec:discussion}
In the context of a QRNG, the main result of the phase diffusion measurements are dependences $\sigma_{\varphi}(I_b^{exp})$ and $\sigma_{s}(I_b^{exp})$ shown in Fig.~\ref{fig:sigmaPhiSigmas}. These dependences clearly show what $I_b^{exp}$ values should be used to ensure that interference of laser pulses acts as uncorrelated quantum entropy source. We see from Fig.~\ref{fig:sigmaPhiSigmas} that for 1.25~GHz the bias current values $I_b^{exp}>43$~mA yield $\sigma_\varphi<\pi$. According to Eq.~\eqref{eq:QRF}, this means that the phase between neighboring laser pulses correlates significantly, such that these pulses cannot be used for QRNG. In the range between 43 and 37~mA, $\sigma_\varphi$ belongs to the range $[\pi,2\pi]$, such that we should use the factor $\tilde{\Gamma}$ (Eq.~\eqref{eq:gammaTilde}) for randomness extraction. Finally, when $I_b^{exp}<37$~mA, we may neglect any phase correlations between laser pulses and assume that $f_{\Delta\varphi}$ is uniform, which allows employing the factor $\Gamma$ from Eq.~\eqref{eq:quantumReductionFactor}. At 2.5~GHz, the laser should be employed at values $I_b^{exp}<28$~mA: between 19 and 28~mA we should use the QRF $\tilde{\Gamma}$, whereas at $I_b^{exp}<19$~mA one may use $\Gamma$.

Equally important are dependencies $\sigma_s(I_b^{exp})$. As was shown in \cite{Shakhovoy2020}, experimental estimation of the QRF is hampered by the fact that the definition of the min-entropy (Eq.~\eqref{eq:minEntropy}) contains the quantity $\tilde{S}_{min}$, which is difficult to determine in practice. It was therefore proposed to determine $\Gamma$ via precalculated curves $\Gamma(B)$, where $B$ is defined by the following ratio:
\begin{equation}
B=\frac{\text{whole PDF width}}{\text{distance between PDF maxima}}.
\end{equation}
The curve $\Gamma(B)$ can be computed with the use of Eqs.~\eqref{eq:resultOfInterference8} and \eqref{eq:uniformPDF} in assumption of a certain value of $\sigma_s$. Experimentally measured dependencies $\sigma_s(I_b^{exp})$ thus allow computing proper dependencies $\Gamma(B)$.

The nature of the damped oscillations in theoretical curves $\sigma_\varphi(I_b)$ in Fig.~\ref{fig:CUDAsimulations} can be understood by referring to the work of C.~Henry \cite{Henry83}, where the author provides an explanation for the additional peaks appearing in a spectrum of a semiconductor laser. The satellite peaks were related to the time dependence of the mean square phase change  $\langle\Delta\varphi^2\rangle$, which, in addition to a linear dependence, exhibits damped periodical variations caused by relaxation oscillations. It was shown that
\begin{equation}\label{eq:meanSquarePhaseChange}
\langle\Delta\varphi^2\rangle\propto\exp(-\gamma_d t)\cos(\omega_r t-3\delta),
\end{equation}
where $\gamma_d$  and $\omega_r$  are the damping rate and the angular frequency of relaxation oscillations, respectively, and  ${\cos\delta=\omega_r/\sqrt{\omega_r^2+\gamma_d^2}}$. Obviously, the quantity $\sigma_\varphi$  we have considered so far corresponds to $\sqrt{\langle\Delta\varphi^2\rangle}$; however, the mean square phase change  $\langle\Delta\varphi^2\rangle$ in Eq.~\eqref{eq:meanSquarePhaseChange} was derived for the continuous wave lasing (i.e., above threshold), whereas the standard deviation plotted in Fig.~\ref{fig:CUDAsimulations} was calculated for the "large signal". In other words, $\sigma_\varphi$ in Fig.~\ref{fig:CUDAsimulations} contains the evolution of the phase below threshold (when the laser does not emit) as well as above threshold (when the laser emits the pulse). We may thus divide $\sigma_\varphi$ into two parts: the one part is related to the phase evolution below threshold and should decrease monotonically with the increase of $I_b$, whereas the second part is related to the phase diffusion above threshold (during the laser pulse emission) and should exhibit oscillations according to Eq.~\eqref{eq:meanSquarePhaseChange}. In fact, increase of the bias current leads to the increase of the laser pulse duration, which is equivalent to the increase of time $t$ in Eq.~\eqref{eq:meanSquarePhaseChange}. So, damped oscillation in the $\sigma_\varphi(I_b)$ curves are driven by the competition between the oscillating growth of $\langle\Delta\varphi^2\rangle$ with $I_b$ during laser pulse emission and monotonic decrease of $\langle\Delta\varphi^2\rangle$ with $I_b$ between laser pulses. At sufficiently high values of $I_b$, when the width of the laser pulse becomes approximately equal to the width of the electric pulse and thus ceases to grow with increasing $I_b$, the dependence $\sigma_\varphi(I_b)$ becomes monotonically decreasing.

\section{Conclusions}\label{sec:conclusion}
In this work, we provided a general description of a quantum entropy source based on the interference of laser pulses and explicitly formulated important problems (of both a purely theoretical and practical nature), which are often overlooked in the development of QRNGs. We have considered in detail the different modes of laser pulse interference and have shown that the interference signal remains quantum in nature even in the presence of classical phase drift in the interferometer if the phase diffusion obeys the relation $\sigma_\varphi>2\pi$. We have explicitly formulated the relationship between the previously introduced quantum reduction factor \cite{Shakhovoy2020} and the leftover hash lemma. Using this relationship, we developed a method to estimate the quantum noise contribution to the interference signal in the presence of phase correlations when $\pi<\sigma_\varphi<2\pi$. Finally, in addition to conventional interference fringes, we proposed to use statistical interference fringes to obtain more detailed information about the probabilistic properties of laser pulse interference. We hope that the theoretical and experimental results presented here will be useful for the developers of QRNGs, and may also help in the preparation of probabilistic models necessary for the certification of such devices.

\begin{acknowledgments}
We are grateful to A. Losev and V. Zavodilenko for assistance in the development of the laser driver and to A. Udaltsov for the development of the TOPM-controller.
\end{acknowledgments}

\appendix
\section{Derivation of stochastic rate equations}\label{sec:derivationOfStochactisEquations}
According to C. Henry \cite{Henry82}, a spontaneous emission event is described as a randomly occurring increase of the field amplitude and an accompanying random change in the phase $\varphi$  of the optical field. The complex field amplitude $E(t)$  is assumed to increase by  $\Delta {E_k}$, which has an absolute value equal to 1 and the phase  $\varphi  + {\theta _k}$, i.e.,  $\Delta {E_k} = \exp [i(\varphi  + {\theta _k})]$, where $\theta _k$  is a random angle. Note that the electric field $E$  is assumed here to be normalized such that its absolute square corresponds to the average photon number $Q$ inside the laser cavity: $Q = {\left| E \right|^2}$  or  $E = \sqrt Q \exp (i\varphi )$. It is important to emphasize that despite its correspondence to the average photon number, the quantity   should be referred to as a \textit{normalized intensity} rather than to as a photon number,  $N_{ph}$. This feature is not relevant when considering dynamics of averaged quantities since $\left\langle Q \right\rangle  = \left\langle {{N_{ph}}} \right\rangle$  (here, angular brackets denote ensemble averaging); however, it becomes relevant when considering stochastic equations, inasmuch as $N_{ph}$  and $Q$  have different distributions and diffusion coefficients \cite{Henry86}.

It is assumed that different events producing the change $\Delta {E_k}$  are uncorrelated, such that the spontaneous emission noise $F_E(t)$  may be written as a sequence of $\delta$-pulses:
\begin{equation}\label{eq:FEasDeltaFunctions}
{F_E}(t) = \sum\limits_k {\Delta {E_k}} \delta (t - {t_k})
\end{equation}
with uncorrelated  $t_k$. The rate of these events obviously corresponds to the average rate of radiative spontaneous emission into the lasing mode, which can be written as  $C_{sp}R_{sp}$, where  $R_{sp}$ is the total average rate of radiative spontaneous emission and the $C_{sp}$  factor corresponds to the fraction of spontaneously emitted photons that end up in the active mode under consideration. According to a general approach, the spontaneous emission noise may be introduced just by adding the complex Langevin force  ${F_E}(t)$ to the right-hand side of the single-mode laser rate equation for the complex slowly varying electric field amplitude \cite{Petermann,Henry82,Henry86,Tartwijk95,AgrawalDutta}:
\begin{equation}\label{eq:fieldRateEquation}
dE = \frac{1}{{2{\tau _{ph}}}}(1 + i\alpha )({G_L} - 1)Edt + {F_E}(t)dt,
\end{equation}
where $\alpha$  is the linewidth enhancement factor (the Henry factor \cite{Henry82}), $\tau_{ph}$  corresponds to the inverse decay rate of the field intensity and is generally treated as a photon lifetime, and the normalized dimensionless linear gain ${G_L} = {{(N - {N_{tr}})}/{({N_{th}} - {N_{tr}})}}$  depends on the carrier number $N$, where  $N_{tr}$ and $N_{th}$ are the carrier numbers at transparency and threshold, respectively. According to Eq.~\eqref{eq:FEasDeltaFunctions}, $F_E(t)$  satisfies the general relations:
\begin{equation}
\left\langle {{F_E}(t)} \right\rangle  = 0,\,\,\,\,\left\langle {{F_E}(t){F_E}(t - \tau )} \right\rangle  = 0,
\end{equation}
and
\begin{equation}\label{eq:FEautocorrlation}
\left\langle {{F_E}(t)F_E^*(t - \tau )} \right\rangle  = 2{D_{EE}}\delta (\tau ),
\end{equation}
where the asterisk means the complex conjugate. To determine the coefficient $D_{EE}$  one should compute  $\left\langle {{F_E}(t)F_E^*(u)} \right\rangle$, where  $u = t - \tau$. According to Eq.~\eqref{eq:FEasDeltaFunctions}, this quantity is the product of two sums of delta functions, where all cross terms are zero since $\delta (t - {t_k})\delta (u - {t_l})$  is zero unless  ${t_k} = {t_l}$; hence
\begin{equation}\label{eq:productOfFE}
{F_E}(t)F_E^*(u) = \sum\limits_k {\delta (t - u)\delta (u - {t_k})} .
\end{equation}
Finally, the averaging can be performed by replacing ${\sum _k}$  in Eq.~\eqref{eq:productOfFE} by ${C_{sp}}{R_{sp}}\int {d{t_k}}$. Thus, we have after averaging:
\begin{equation}
\left\langle {{F_E}(t)F_E^*(t - \tau )} \right\rangle  = {C_{sp}}{R_{sp}}\delta (\tau ).
\end{equation}
Comparing this to Eq.~\eqref{eq:FEautocorrlation} we obtain  $2{D_{EE}} = {C_{sp}}{R_{sp}}$, such that we can write
\begin{equation}
{F_E} = \sqrt {\frac{{{C_{sp}}{R_{sp}}}}{2}} \left( {{\xi _1} + i{\xi _2}} \right),
\end{equation}
where  $\xi_1$ and $\xi_2$  are independent random variables representing the normalized white Gaussian noise and obey the following relations:
\begin{equation}
\langle {\xi _i}(t)\rangle  = 0,\,\,\,\,\langle {\xi _i}(t){\xi _j}(t - \tau )\rangle  = {\delta _{ij}}\delta (\tau ).
\end{equation}

Writing the normalized electric field as $E = {x_1} + i{x_2}$  and separating the real and imaginary parts of Eq.~\eqref{eq:fieldRateEquation} we will obtain the following system:
\begin{equation}
\begin{split}
d{x_1} &= {h_1}dt + {g_{11}}{\xi _1}dt + {g_{12}}{\xi _2}dt,\\
d{x_2} &= {h_2}dt + {g_{21}}{\xi _1}dt + {g_{22}}{\xi _2}dt,
\end{split}
\end{equation}
where
\begin{equation}
\begin{split}
{h_1} &= \frac{1}{{2{\tau _{ph}}}}({G_L} - 1){x_1} - \frac{\alpha }{{2{\tau _{ph}}}}({G_L} - 1){x_2},\\
{h_2} &= \frac{1}{{2{\tau _{ph}}}}({G_L} - 1){x_2} + \frac{\alpha }{{2{\tau _{ph}}}}({G_L} - 1){x_1},\\
{g_{11}} &= {g_{22}} = \sqrt {\frac{{{C_{sp}}{R_{sp}}}}{2}} ,\,\,\,\,{g_{12}} = {g_{21}} = 0.
\end{split}
\end{equation}
Introducing variables $x_1'$  and $x_2'$  by the relations
\begin{equation}
\begin{split}
{x_1'} &= {u_1}({x_1},{x_2}) \equiv Q = x_1^2 + x_2^2,\\
{x_2'} &= {u_2}({x_1},{x_2}) \equiv \varphi  = \arctan ({x_2/x_1}),
\end{split}
\end{equation}
and using the It\^{o} formula \cite{Kloeden} written for our case as
\begin{widetext}
	\begin{equation}\label{eq:itoFormula}
	d{{x'}_i} = \left\{ {\frac{{\partial {u_i}}}{{\partial t}} + \sum\limits_{k = 1}^2 {{h_k}\frac{{\partial {u_i}}}{{\partial {x_k}}}}  + \frac{1}{2}\sum\limits_{j = 1}^2 {\sum\limits_{k,m = 1}^2 {{g_{mj}}{g_{kj}}\frac{{{\partial ^2}{u_i}}}{{\partial {x_m}\partial {x_k}}}} } } \right\}dt
	+ \sum\limits_{j = 1}^2 {\sum\limits_{k = 1}^2 {{g_{kj}}\frac{{\partial {u_i}}}{{\partial {x_k}}}} } \,d{W_j},
	\end{equation}
\end{widetext}
we will obtain the following set of stochastic rate equations:
\begin{equation}\label{eq:fieldRateEquationInDifferentials}
\begin{split}
dQ &= ({G_L} - 1)\frac{Q}{{{\tau _{ph}}}}dt + {C_{sp}}{R_{sp}}dt + {F_Q}dt,\\
d\varphi & =   \frac{\alpha }{{2{\tau _{ph}}}}({G_L} - 1)dt + {F_\varphi }dt
\end{split}
\end{equation}
with
\begin{equation}\label{eq:stochasticTerms}
\begin{split}
{F_Q}dt &= \sqrt {2{C_{sp}}{R_{sp}}Q} \left[ {\cos (\varphi )d{W_1} + \sin (\varphi )d{W_2}} \right],\\
{F_\varphi }dt &= \sqrt {\frac{{{C_{sp}}{R_{sp}}}}{{2Q}}} \left[ {\cos (\varphi )d{W_2} - \sin (\varphi )d{W_1}} \right],
\end{split}
\end{equation}
where  $W_1$, $W_2$ are two independent Wiener processes, which may be defined via the notation $d{W_1} = {\xi _1}dt$  and  $d{W_2} = {\xi _2}dt$. (It should be remembered, however, that the Wiener process $W(t)$  is nowhere differentiable, so the equality $dW(t) = \xi (t)dt$  cannot be treated as a differential in the usual sense and it is better to consider it just as a symbolic notation \cite{Kloeden}.)

Since the gain  $G_L$ depends on the carrier number  $N$, Eqs.~\eqref{eq:fieldRateEquationInDifferentials} should be supplemented by the rate equation for  $N$. If one can neglect carrier diffusion effects, as, e.g., in index-guided lasers, the rate equation for the carrier number can be written in the following form:
\begin{equation}\label{eq:rateEquationForN}
dN = \frac{I}{e}dt - \frac{N}{{{\tau _e}}}dt - \frac{{{G_L}Q}}{{\Gamma {\tau _{ph}}}}dt + {F_N}dt,
\end{equation}
where  $I$ is the pump current, $e$  is the absolute value of the electron charge,  $\tau_e$ is the effective carrier lifetime, and $\Gamma$  is the confinement factor, which can be estimated as a ratio between the cross-sectional areas of the transverse mode and the active layer. We also added in Eq.~\eqref{eq:rateEquationForN} the Langevin force  $F_N$, which drives fluctuations of  $N$; however, due to relatively short ($\sim1$~ns) carrier lifetime, perturbations to the carrier density do not persist long enough to make significant low-frequency contributions, while at higher frequencies they are damped by diffusion \cite{Lang85}. Therefore, carrier fluctuations are usually assumed to be negligible, when modeling stochastic properties of laser radiation. We will thus assume below that  $F_N=0$.

Finally, rate equations for $N$ and $Q$ should be modified to take into account the gain saturation \cite{Agrawal88,Agrawal90}. This can be performed by substituting $G_L$  by  $G = {{{G_L}}/{\sqrt {1 + 2{\gamma _Q}Q} }} \approx {G_L}(1 - {\gamma _Q}Q)$, where $\gamma_Q$  is the dimensionless gain compression factor. (The rate equation for the phase is generally left unchanged.) Sometimes, it is more convenient to use instead of $\gamma_Q$  the gain compression factor $\gamma_P$  related to the output optical power $P$, which can be defined as  $\gamma_P  = {{2{\gamma_Q}\Gamma {\tau _{ph}}}/{(\eta_d \hbar {\omega _0})}}$, where $\eta_d$  is the differential quantum output (don't confuse it with visibility, for which we used above similar notation), and  $\hbar\omega_0$ is the photon energy. For typical semiconductor lasers, $\gamma_P$  may reach a few dozens of W$^{-1}$ \cite{Shakhovoy2021}; here we will assume that it is in the range 10--40~W$^{-1}$. 

The system of stochastic differential equations (SDEs) \eqref{eq:fieldRateEquationInDifferentials} and \eqref{eq:rateEquationForN} with stochastic terms given by Eq.~\eqref{eq:stochasticTerms} can be solved numerically with the Euler-Maruyama method \cite{Kloeden}, which is the simplest time discrete approximation used for integration of SDEs. In this method, each component of the solution of the vector valued SDE is approximated by a continuous time stochastic process defined via the iterative scheme with the time differential $dt$  substituted by the conventional time increment $\Delta  = {t_{n + 1}} - {t_n}$  and the independent "differentials"  $dW_1$ and $dW_2$ approximated by the increments  $\Delta {W_{1,2}} = {W_{1,2}}({t_{n + 1}}) - {W_{1,2}}({t_n})$, which are assumed to be normally distributed with zero mean value and variance equal to  $\Delta$. We can write  $\Delta {W_{1,2}} = {\xi _{1,2}}\sqrt \Delta $, where $\xi_{1,2}$  are discrete Gaussian random variables with zero mean and variance equal to 1. Thereby, our system of SDEs can be written in the following form suitable for the direct numerical integration:
\begin{widetext}
	\begin{equation}\label{eq:differenceSheme}
	\begin{split}
	{Q_{n + 1}} &= {Q_n} + \left( {\frac{{{N_n} - {N_{tr}}}}{{{N_{th}} - {N_{tr}}}}\frac{1}{{\sqrt {1 + 2{\gamma _Q}{Q_n}} }} - 1} \right)\frac{{{Q_n}}}{{{\tau _{ph}}}}\Delta  + {C_{sp}}\frac{{{N_n}}}{{{\tau _e}}}
	+ 2\sqrt {\frac{{{C_{sp}}{N_n}{Q_n}}}{{2{\tau _e}}}} \left( {\xi _1^n\cos {\varphi _n} + \xi _2^n\sin {\varphi _n}} \right)\sqrt \Delta  ,\\
	{\varphi _{n + 1}}& = {\varphi _n} + \frac{\alpha }{{2{\tau _{ph}}}}\left( {\frac{{{N_n} - {N_{tr}}}}{{{N_{th}} - {N_{tr}}}} - 1} \right)\Delta 
	+ \sqrt {\frac{{{C_{sp}}{N_n}}}{{2{\tau _e}{Q_n}}}} \left( {\xi _2^n\cos {\varphi _n} - \xi _1^n\sin {\varphi _n}} \right)\sqrt \Delta  ,\\
	{N_{n + 1}} &= {N_n} + \frac{{{I_n}}}{e}\Delta  - \frac{{{N_n}}}{{{\tau _e}}}\Delta  - \frac{{{Q_n}}}{{\Gamma {\tau _{ph}}}}\frac{{{N_n} - {N_{tr}}}}{{{N_{th}} - {N_{tr}}}}\frac{1}{{\sqrt {1 + 2{\gamma _Q}{Q_n}} }}\Delta ,
	\end{split}
	\end{equation}
\end{widetext}
where we used the notations:  ${Q_n} \equiv Q({t_n})$,  ${N_n} \equiv N({t_n})$,  ${\varphi _n} \equiv \varphi ({t_n})$,  ${I_n} \equiv I({t_n})$,  $\xi _{1,2}^n \equiv {\xi _{1,2}}({t_n})$.

It should be taken into account that direct implementation of the Euler-Maruyama scheme may lead to an unphysical result, namely to negative values of  $N$ and  $Q$; therefore, Eq.~\eqref{eq:differenceSheme} should be solved with constraint that  $N$ and $Q$  are non-negative. Note that this feature persists for the second order scheme (e.g., the Milstein scheme \cite{Kloeden}) albeit the latter somewhat minimizes such unphysical trajectories. In simulations shown below, we used the first order (Euler-Maruyama) scheme. Note that we performed integration at different reasonable integration steps  $\Delta  < 0.1{\tau _{ph}}$; the results obtained at different $\Delta$  were the same for all simulations.

\section{Controlling the interferometers}\label{app:interferometerControl}
As was discribed in section \ref{sec:experimentalSetup}, we used for measurements a chip with a cascade of unbalanced Mach-Zehnder interferometers (two of them are shown in Fig.~\ref{fig:uMZI}). To choose the desired delay line, we used an additional balanced interferometers placed between the unbalanced ones; there were seven interferometers on the chip in total: 4 balanced and 3 unbalanced with the delay lines $\Delta T_1=200$~ps, $\Delta T_2=400$~ps, and $\Delta T_3=800$~ps (each MZI was also equipped by a thermo-optical phase modulator). Due to such a configuration, the evolution of phase in the interferometers have a cumulative effect, such that splitting ratios depend on the set of phases in all four balanced interferometers. To determine the phases that must be set on thermo-optical modulators in order to select the required delay line, let us write a system of recursive equations that describe propagation of an optical pulse through a cascade of interferometers:
\begin{equation}\label{eq:fieldsInInterferometers}
\begin{split}
E_{12}^\pm(t)&=1/4\cdot E_{in}(t) (1\pm e^{i\varphi_c^1}),\\
E_{23}^\pm(t)&=1/\sqrt{2}\cdot[E_{12}^-(t-\Delta T_1)\pm E_{12}^+(t)e^{i\varphi_{u}^1}],\\
E_{34}^\pm(t)&=1/\sqrt{2}\cdot[E_{23}^-(t)\pm E_{23}^+(t)e^{i\varphi_c^2}],\\
E_{45}^\pm(t)&=1/\sqrt{2}\cdot[E_{34}^-(t-\Delta T_2)\pm E_{34}^+(t)e^{i\varphi_{u}^2}],\\
E_{56}^\pm(t)&=1/\sqrt{2}\cdot[E_{45}^-(t)\pm E_{45}^+(t)e^{i\varphi_c^3}],\\
E_{67}^\pm(t)&=1/\sqrt{2}\cdot[E_{56}^-(t-\Delta T_3)\pm E_{56}^+(t)e^{i\varphi_{u}^3}],\\
E_{out}^\pm(t)&=1/\sqrt{2}\cdot[E_{67}^-(t)\pm E_{67}^+(t)e^{i\varphi_c^4}],
\end{split}
\end{equation}
where $E_{in}(t)$ is the electric field in the input laser pulse, $E_{mn}^{\pm}(t)$ stands for the electric field in the laser pulse being between the $m$-th and $n$-th interferometers at time $t$, and the signs "$+$" and "$-$" correspond to different ports of the corresponding directional coupler. Photodetectors connected to the output ports $+$ and $-$ measure the intensities $|E_{out}^+|$ and $|E_{out}^-|$, respectively. (We will assume for simplicity that $|E_{in}(t)|^2$ is defined by the Gaussian curve: $|E_{in}(t)|^2=\exp[-t^2/(2w^2)]$, where $w$ is the rms width of the pulse.) 

\begin{figure}[b]
	\includegraphics[width=\columnwidth]{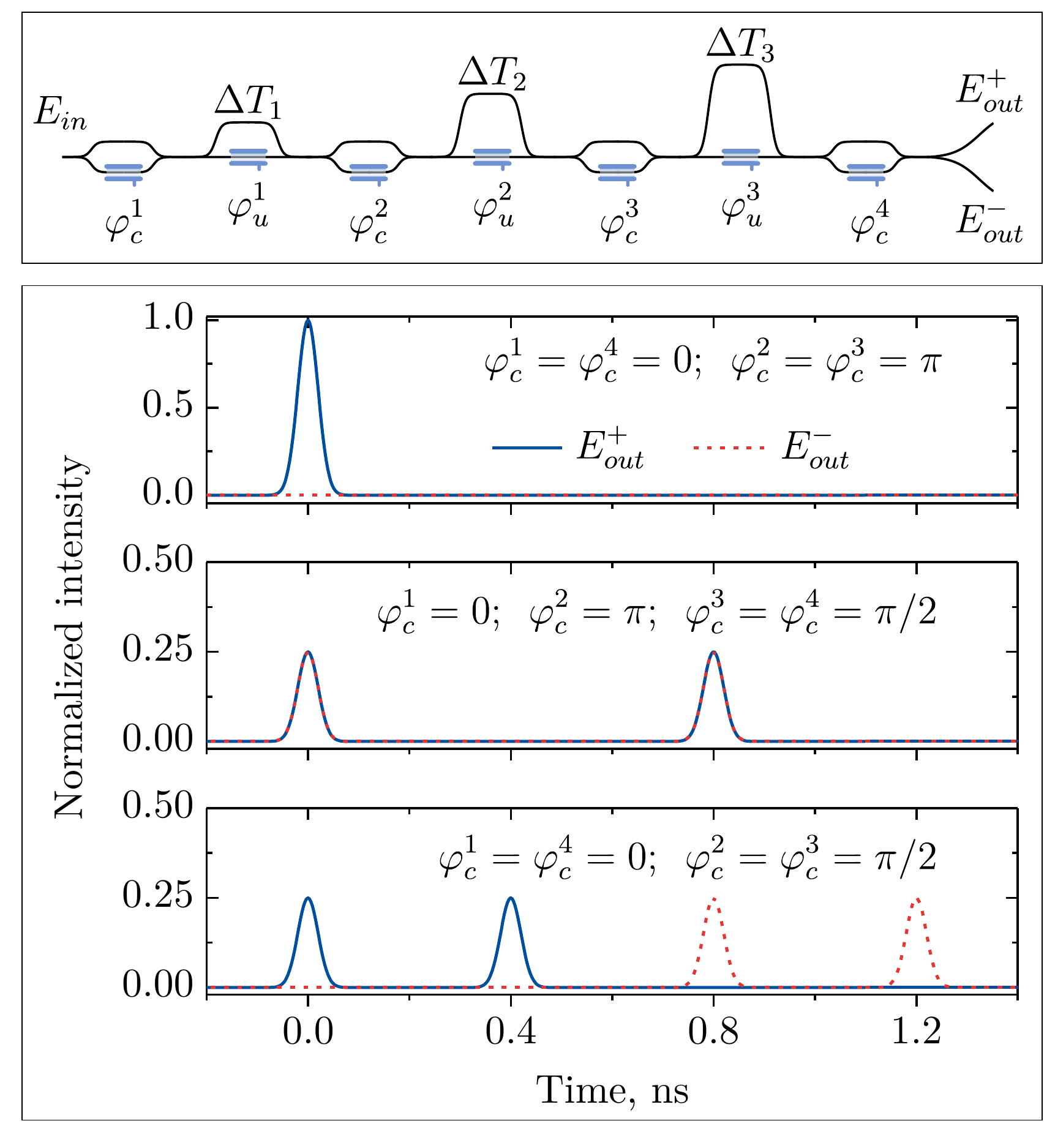}
	\caption{\label{fig:choiceOfTheDelayLine} Schematic representation of a cascade of interferometers on the chip we used in the experiment (on the top) and the output of the system for various configurations of phases $\varphi_c^k$ in assumption that a single laser pulse of the Gaussian shape with the rms width of 20~ps arrives at the input of the interferometer.}
\end{figure}

It is easy to see that configuration of splitting ratios, which defines the choice of the delay line, does not depend on phases $\varphi_u^k$ determining phase shifts in unbalanced interferometers; therefore, for simplicity, they can be put to zero. It is interesting to note that if we put also $\varphi_c^1=\varphi_c^2=\varphi_c^3=\varphi_c^4=0$, we will obtain using \eqref{eq:fieldsInInterferometers}:
\begin{equation}
|E_{out}^-|=0,\,\,\,\,|E_{out}^+|=\exp\left[-\frac{(t-\Delta T_2)^2}{2w^2}\right],
\end{equation}
i.e., the incoming laser pulse passes through the delay line $\Delta T_2=400$~ps and outputs to the port "$+$" (and nothing is output to the port "$-$"). 

To "close" all the delay lines, the following phase configuration should be used:  $\varphi_c^1=\varphi_c^4=0$,  $\varphi_c^2=\varphi_c^3=\pi$, which yields
\begin{equation}\label{eq:delaLineClose}
|E_{out}^-|=0,\,\,\,\,|E_{out}^+|=\exp\left[-t^2/(2w^2)\right].
\end{equation}
To choose the interferometer with $\Delta T_3=800$~ps, one should put: $\varphi_c^1=0$, $\varphi_c^2=\pi$, $\varphi_c^3=\varphi_c^4=\pi/2$, which provides
\begin{equation}\label{eq:delayLine800}
|E_{out}^\pm|^2=\frac{1}{4}\left[e^{-\frac{t^2}{4w^2}}+e^{-\frac{(t-\Delta T_3)^2}{4w^2}}\right]^2.
\end{equation}
To choose the interferometer with $\Delta T_2=400$~ps, one may use:
$\varphi_c^1=\varphi_c^4=0$,  $\varphi_c^2=\varphi_c^3=\pi/2$, which gives
\begin{equation}\label{eq:delayLine400}
\begin{split}
|E_{out}^+|^2&=\frac{1}{4}\left[e^{-\frac{t^2}{4w^2}}+e^{-\frac{(t-\Delta T_2)^2}{4w^2}}\right]^2,\\
|E_{out}^-|^2&=\frac{1}{4}\left[e^{-\frac{(t-\Delta T_3)^2}{4w^2}}+e^{-\frac{(t-\Delta T_2-\Delta T_3)^2}{4w^2}}\right]^2.
\end{split}
\end{equation}
The results corresponding to Eqs.~\eqref{eq:delaLineClose}--\eqref{eq:delayLine400} are shown in Fig.~\ref{fig:choiceOfTheDelayLine} in assumption that a single laser pulse of the Gaussian shape with the rms width of 20~ps arrives at the input of the interferometer.

\section{Signal normalization}\label{app:signalNormalization}
\begin{figure}[t]
	\includegraphics[width=0.95\columnwidth]{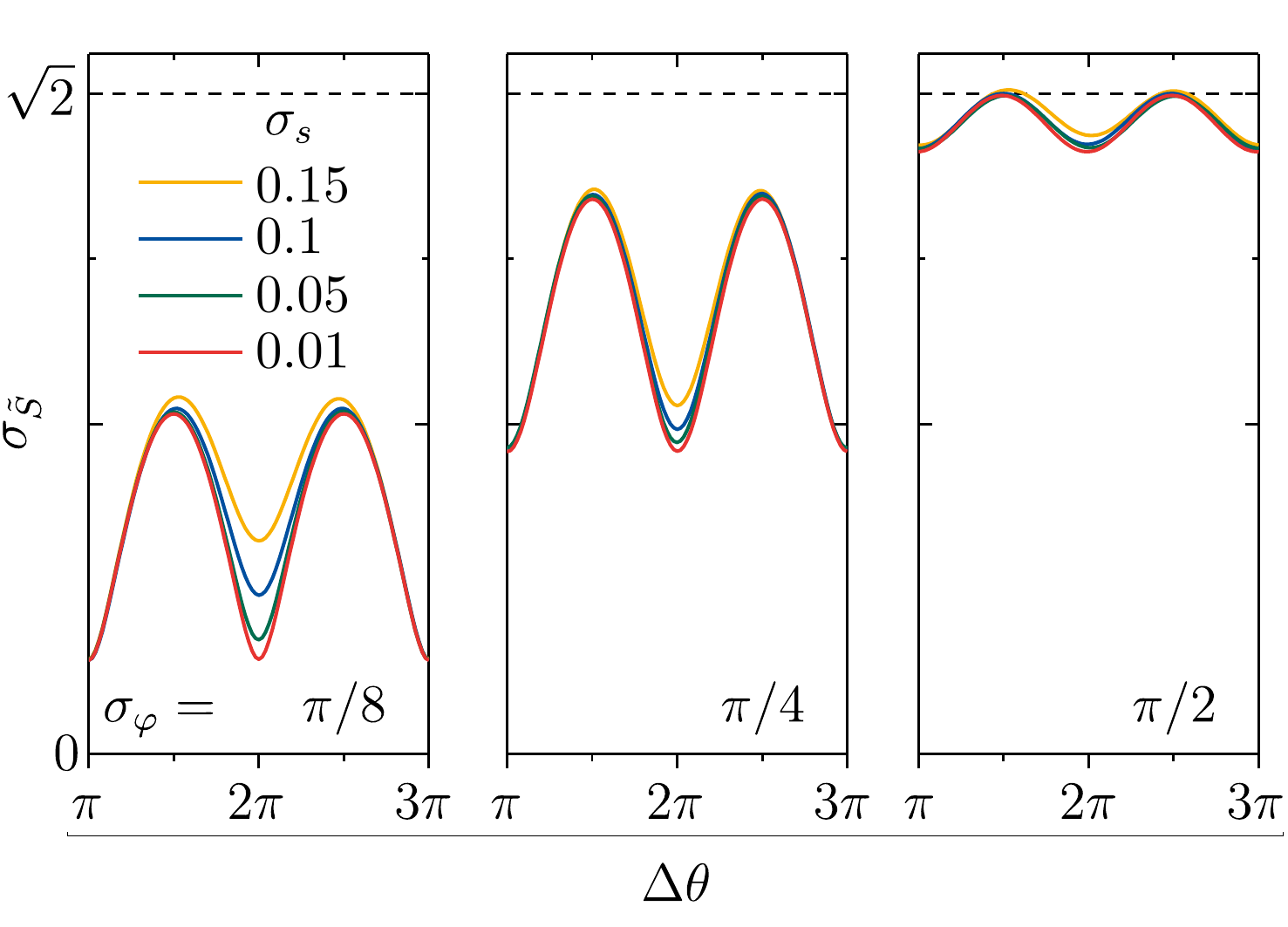}
	\caption{\label{fig:statisticalFringesTheoretical} Theoretical statistical fringes at various values of $\sigma_\varphi$ and $\sigma_s$.}
\end{figure}
When constructing experimental PDFs, it is extremely important to perform proper normalization. To obtain the normalized signal defined by Eq.~\eqref{eq:normalizedSignal}, it was taken into account that even in the absence of optical power, the "area" under the signal from the photodetector (i.e. the measured energy contained in the pulse) can be different from zero; therefore, in all experiments, we preliminary determined the "origin" $S_{zero}$. In addition, we estimated the losses in the interferometer arm. For this, we set the pulse repetition rate to 312.5~MHz (the repetition period of 3.2~ns) and set the voltage on the TOPM-controller such that the optical pulse does not enter any of the delay lines, and all the corresponding optical power goes out to the port connected to the photodetector. The optical energy per pulse was determined as the area under the photodetector signal (in picowebers) in the range from 0 to 3.2~ns (the position of zero on the time axis was chosen arbitrarily). ($S_{zero}$ was measured as the area of the photodetector signal in the same range, but with the laser turned off.) Then the delay line of 800 ps was selected and the area under the photodetector signal was again measured in the same range, the value of which we designated now $S_{800}$. It is clear from Eq.~\eqref{eq:delayLine800} (see also Fig.~\ref{fig:choiceOfTheDelayLine}) that the insertion loss of the delay line $\Delta T_3=800$~ps (in dB) can be estimated using the formula ${\alpha _{800}} = 10\,{\log _{10}}\{ {{({S_0} - {S_{zero}})}/{[2({S_{800}} - {S_{zero}})]}}\}$. We obtained $\alpha_{800}\approx 1.3$~dB. The pulse repetition rate was then set to 1.25~GHz and the pulse energy $S_0$ (without entering the delay line) was measured in the range from 0 to 800~ps. Then, the histogram of the interference PDF (over the pulse area) was recorded. Each value along the $x$-axis was normalized according to the formula:
\begin{equation}
\tilde S = {10^{\alpha_{800}/10}}\frac{{x - {S_{zero}}}}{{{S_0} - {S_{zero}}}}.
\end{equation}
In a similar way, we normalized he signal at 2.5~GHz. For the delay line $\Delta T_2=400$~ps we obtained $\alpha_{400}\approx0.7$~dB.

\section{$\sigma_{\tilde{S}}$-fringes}\label{app:statisticalFringes}
Figure \ref{fig:statisticalFringesTheoretical} demonstrates theoretical statistical interference fringes at various values of $\sigma_\varphi$ and $\sigma_s$. One can see that the depth of the dip in the $\sigma_{\tilde{S}}$-curve at constructive interference ($\Delta\theta=0$) decreases when increasing $\sigma_s$, and such an asymmetry of the fringe is more pronounced at smaller $\sigma_\varphi$. Although the analysis of $\sigma_{\tilde{S}}$-fringes is not as straightforward as fitting conventional interference fringes, it allows getting deeper insight into statistical features of the pulse interference and can be thus a useful tool.


\bibliography{phaseRandomnessRefs}

\end{document}